\documentclass[prb,aps,reprint, twocolumn,showpacs,amsmath,superscriptaddress,longbibliography,notitlepage]{revtex4-2}
\usepackage{amssymb,mathtools,amsmath}
\usepackage{physics}
\usepackage[utf8]{inputenc}
\usepackage{mathrsfs}
\usepackage{graphicx}
\usepackage{float}
\usepackage[caption=false]{subfig}
\usepackage{dcolumn}
\usepackage{appendix}
\usepackage{tabularx}
\usepackage{epstopdf}
\usepackage[normalem]{ulem}
\usepackage{verbatim}
\usepackage{bm}
\usepackage{natbib}
\usepackage{multirow}
\usepackage{tikz}
\usepackage{empheq}
\usetikzlibrary{fit}
\usepackage{footnote}
\usepackage{nicematrix}
\usepackage{xcolor}
\usepackage{hyperref}
\hypersetup{
	colorlinks,
	citecolor=violet,
	linkcolor=red,
	urlcolor=violet}
	
\usepackage{orcidlink}
\usepackage{nicematrix}

\DeclareMathAlphabet\mathbfcal{OMS}{cmsy}{b}{n}

\usepackage{soul}
\begin{document}

\title{Topolectrical circuits -- recent experimental advances and developments}

\author{Haydar Sahin\,\orcidlink{0000-0002-6361-8161}}
\email{sahinhaydar@u.nus.edu}
\affiliation{Department of Electrical and Computer Engineering, National University of Singapore, Singapore 117583, Republic of Singapore}

\author{Mansoor B.~A. Jalil\,\orcidlink{0000-0002-9513-8680}}
\email{elembaj@nus.edu.sg}
\affiliation{Department of Electrical and Computer Engineering, National University of Singapore, Singapore 117583, Republic of Singapore}

\author{Ching Hua Lee\,\orcidlink{0000-0003-0690-3238}}
\email{phylch@nus.edu.sg}
\affiliation{Department of Physics, National University of Singapore, Singapore 117542, Republic of Singapore}

\begin{abstract}
Metamaterials serve as versatile platforms for demonstrating condensed matter physics and non-equilibrium phenomena, with electrical circuits emerging as a particularly compelling medium. This review highlights recent advances in the experimental circuit realizations of topological, non-Hermitian, non-linear, Floquet and other notable phenomena. Initially performed mostly with passive electrical components, topolectrical circuits have evolved to incorporate active elements such as operational amplifiers and analog multipliers that combine to form negative impedance converters, complex phase elements, high-frequency temporal modulators and self-feedback mechanisms. This review provides a summary of these contemporary studies and discusses the broader potential of electrical circuits in physics.
\end{abstract}

\maketitle

{\hypersetup{linkcolor=black}
	\tableofcontents}

\section{Introduction}
The use of electrical circuit metamaterial arrays to simulate condensed matter phenomena has emerged as a popular approach alongside other metamaterial approaches~\cite{ni_topological_2023} such as mechanical~\cite{nash_topological_2015,zheng_progress_2022,wang_non-hermitian_2023}, acoustical~\cite{ma_topological_2019,xue_topological_2022,xue_realization_2019,he_hybrid_2019,ding_experimental_2019,wei_higher-order_2021}, and photonic~\cite{khanikaev_photonic_2013,lu_topological_2014,ozawa_topological_2019} media. Numerous studies, some of which are reviewed in this article, have demonstrated that a wide range of condensed matter phenomena can be experimentally realized through the design of appropriate electrical circuits. The use of electrical circuit arrays to study the resistance or conductance between the farthest points of a lattice continuum has a long history in condensed matter circles~\cite{aitchison_resistance_1964,bartis_lets_1967,kirkpatrick_percolation_1973,cserti_application_2000,wu_theory_2004,tzeng_theory_2006} and has been studied through various approaches such as lattice Green's function~\cite{cserti_uniform_2011,mamode_laplacian_2024}, transfer matrix method~\cite{perrier_leading_2024}, recursion–transform approach~\cite{tan_recursion-transform_2015}, and method of images~\cite{mamode_electrical_2017}. However, the formal connection between the Hamiltonian formalism and the circuit Laplacian was established less than a decade ago. Early work on periodic circuit lattices primarily addressed point-to-point resistance or impedance problems, lacking a comprehensive methodological framework that connect with condensed matter models. 

One of the first works that introduced a systematic approach to this was the paper by Lee et al. in 2018~\cite{lee_topolectrical_2018}, which introduced a systematic approach that formalized the concept of Topolectrical (TE) circuits. It is inspired by two earlier studies by Ningyuan et al.~\cite{ningyuan_time-_2015} and Albert et al.~\cite{albert_topological_2015} that provided explicit constructions of circuit lattices exhibiting topological properties~\cite{qi_topological_2011,kane_z_2005,hasan_colloquium_2010,moore_topological_2007,zhang_topological_2009,bernevig_quantum_2006,fu_topological_2007,hasan_three-dimensional_2011,lee2014lattice,schindler_higher-order_2018,ferdous_observation_2023}. Following these foundational works, there has been a rapid proliferation of follow-up works, both through circuit simulations~\cite{helbig_band_2019,hofmann_chiral_2019,haenel_chern_2019,dong_topolectric_2021,liu_topological_2020,yoshida_mirror_2020} and experimental implementations~\cite{imhof_topolectrical-circuit_2018,helbig_generalized_2020,bao_topoelectrical_2019,olekhno_topological_2020,yang_observation_2020,wang_circuit_2020,lu_probing_2019,serra-garcia_observation_2019,yang_observation_2021} (many more to be reviewed later).

Early studies primarily focused on demonstrating topological phenomena~\cite{rafi-ul-islam_system_2022,zhao_topological_2018,li_topological_2018,zhang_topological_2019,liu_octupole_2020,zhang_experimental_2021,luo_topological_2018}, including the realization of edge~\cite{serra-garcia_observation_2019,liu_edge_2021,yatsugi_observation_2022}, higher-order corner~\cite{olekhno_experimental_2022,liu_observation_2022,yang_observation_2022_2,rafi-ul-islam_type-ii_2022,guo_observation_2023}, surface~\cite{zheng_topolectrical_2022} and hinge states~\cite{rafi-ul-islam_chiral_2024}, as well as semi-metalic phases of Weyl~\cite{lu_probing_2019,rafi-ul-islam_topoelectrical_2020,rafi-ul-islam_anti-klein_2020,rafi-ul-islam_realization_2020} or Dirac semi-metals~\cite{yang_experimental_2022,pal_multi_2024}, in multi-dimensional lattices or heterogeneous systems~\cite{sahin_interfacial_2022,sahin_unconventional_2022}. Soon after, these efforts expanded to encompass non-Hermitian~\cite{liu_non-hermitian_2021,ezawa_electric_2019,halder_circuit_2024,su_simulation_2023} and non-linear~\cite{kotwal_active_2021,wang_topologically_2019,zangeneh-nejad_nonlinear_2019} systems, as well as simulators of non-Abelian operators~\cite{wu_non-abelian_2022,pan_three-dimensional_2023} and quantum gates~\cite{ezawa_electric_2020,ezawa_universal_2021} and various other applications~\cite{ezawa2019electric,ezawa2021topological,sahin_topolectrical_2024,yao_multi-orbital_2022,huang_topological_2024,lv_realization_2021,shen2024observation}. At this stage, it is clear that implementing various physical phenomena in electrical circuits has become an engineering task, provided the theoretical framework is validated through the theoretical circuit analysis and circuit simulations.

As such, the focus has recently shifted to the direct device applications of physical models realized by electrical circuits, moving beyond their role as proof-of-concept demonstrations, and the integration of active analog signal processing devices. For the former focus, numerous studies have already proposed sensing applications based on electrical circuits~\cite{zhang_true_2024,yuan_nonhermitian_2023,chen_ultra-sensitivity_2024}. Many of these circuit designs hold promise for practical use, as they can be manufactured using integrated circuit (IC) technologies, enabling their incorporation into everyday technological devices~\cite{rappaport_state_2011,errando-herranz_mems_2020}. For instance, a sensitive boundary-dependent response can be achieved using simple RLC components, thanks to the non-local impedance response of passive electrical circuit components~\cite{zhang_observation_2024}. The latter focus involves the manipulation of signals using devices such as analog multipliers, operational amplifiers (op-amps), and transistors. While op-amps are commonly used to realize non-reciprocal couplings in impedance conversion configurations~\cite{hofmann_chiral_2019,helbig_generalized_2020}, their potential extends far beyond these applications. Circuit configurations involving multiple op-amps and analog multipliers enable to realize numerous phenomena such as Floquet engineering of lattice models~\cite{stegmaier_topological_2024,stegmaier_realizing_2024}, time-varying couplings~\cite{zhang_topolectrical_2025}, or tunable interaction strength~\cite{sun_boundary-localized_2024,zhang_exploring_2024}. One illustrative example is the realization of complex coupling terms using analog multipliers, where four multipliers are configured appropriately to achieve the desired effect~\cite{chen_hyperbolic_2023}.

Below, we highlight some of the draws of using TE circuits:

\begin{itemize}
	\item \textit{Abundant resources}: Since electrical circuits are ubiquitous and well-established, a wealth of resources, ranging from educational to research materials, is readily available in the literature, along with a variety of simulation tools, such as \nobreakdash{LTspice}~\cite{ltspice}, \nobreakdash{PSpice}~\cite{pspice}, \nobreakdash{Multisim}~\cite{multisim} or \nobreakdash{Modelithics~Library}~\cite{modelithics}. The widespread availability of circuit components ensure that the necessary elements are easily accessible. Moreover, these resources are cost-effective, making them suitable for diverse implementations.
	
	\item \textit{Linearity and independence from physical embedding}: Linear circuits can be analyzed using nodal analysis, where each connection at a node represents an additional degree of freedom, analogous to spatial dimensions in physical systems. The wave interactions in these circuits are governed by the principle of linear superposition, allowing the simultaneous synthesis of multiple voltage or current waveforms at a node~\cite{svoboda2013introduction,tooley2007electronic,ulaby_electromagnetics_2005}. Unlike most other metamaterial arrays where propagation direction is constrained by physical geometry, circuits provide the flexibility to emulate systems with higher dimensionalities~\cite{zheng_exploring_2022,wang_circuit_2020}. This capability enables the synthetic realization of effective dimensions that transcend physical constraints, offering a unique advantage over many other platforms.
	
	\item \textit{Straightforward local adjustments and tunability}: Localized modifications, such as defect engineering~\cite{stegmaier_topological_2021}, are straightforward and adaptable. The interconnectness of electrical lattices ensures accessibility to each node, enabling precise local tuning or perturbation~\cite{xie_observation_2024_2,rafi-ul-islam_interfacial_2022,wang_topological_2021}. This offers a significant advantage over less modular setups, where specific localized perturbations are not always possible due to fabrication limitations or intrinsic material properties.

	\item \textit{High reliability and precision}: Well-designed and maintained electrical circuits provide high reliability, even in challenging environmental conditions~\cite{falck_reliability_2018}. They are resilient and do not require special operating environments, such as low temperatures or special electromagnetic shielding~\cite{watts2005high}, making them versatile for various applications~\cite{baker_improved_2014,wang_toward_2013}.
	
	\item \textit{Scalability and accessibility}: Scalability is a major challenge in many metamaterial implementations. Electrical circuits offer high scalability compared to other platforms, which often rely on specialized materials or complex manufacturing techniques~\cite{su_scalability_2023,rumley_silicon_2015}, making scaling up and system modifications more challenging. Additionally, impurities in electrical circuits are typically localized, meaning they do not lead to cumulative effects that cause significant overall systems.
	
	\item \textit{Ease of implementing non-Hermiticity}: Electrical circuits are intrinsically suitable for the realization of non-Hermitian models, which require non-reciprocal couplings or controlled loss~\cite{wu_non-hermitian_2022} and gain~\cite{liu_gain-_2020}. Active components, such as op-amps, allow the direct implementation of highly linear non-reciprocal non-Hermiticity~\cite{liu_non-hermitian_2021}, and loss is naturally present in resistive components~\cite{hofmann_reciprocal_2020}.
	
	\item \textit{Precise control of component uncertainties and loss}: That said, many circuit component manufacturers provide a wide variety of high-quality components with minimal deviations in component parameters and loss. Additionally, for implementations that demand exceptionally low loss and deviation, it is feasible to preselect circuit elements due to their generally low cost, as a variety of low-loss components can be readily found to meet these stringent requirements.
	
	\item \textit{Direct observability and measurability}: Impedance measurements allow key band structure features, such as non-trivial topological zero modes, to be directly observed through straightforward profiling~\cite{imhof_topolectrical-circuit_2018}. Such measurements can either measure the local transmissibility or the global, non-site-specific band structure properties~\cite{franca_impedance_2024}.

\end{itemize}

\noindent Besides these advantages, electrical circuits possess the following set of advantages that are collectively elusive in other platforms:

\begin{itemize}
	\item \textit{Ease of implementing synthetic dimensions}: Independence from physical geometric embedding allows the realization of lattice models in arbitrarily many synthetic dimensions, overcoming physical and spatial limitations~\cite{zhang_topolectrical-circuit_2020,zheng_exploring_2022,wang_circuit_2020,zheng_topological_2024}.

	\item \textit{Broad frequency range compatibility}: Electrical circuits can operate across a wide frequency spectrum, from a few hertz to a few gigahertz~\cite{iizuka_experimental_2023,nagulu_chip-scale_2022,liu_fully_2022}. Unlike photonic systems, which typically require microwave frequencies or higher~\cite{tang_topological_2022}, electrical circuits can function efficiently at much lower AC frequencies (e.g., even a few hundred hertz), making them suitable for studying topological oscillations at timescales amenable to direct human observation. Similarly, while acoustic systems often face challenges in achieving realizations in the ultrasound range (above 20~kHz)~\cite{xue_topological_2022}, integrated circuits can even extend their operation into the terahertz (THz) regime~\cite{sengupta_terahertz_2018}.
	
	\item \textit{Real-time reconfigurability}: The modular nature of circuit components, such as varactors and digitally controlled amplifiers or analog multipliers, enables real-time localized adjustments~\cite{stegmaier_realizing_2024}. This capability is crucial for tasks requiring the dynamic modulation of specific couplings without needing to reconstruct the setup, particularly the simulation of Floquet media~\cite{stegmaier_realizing_2024,stegmaier_topological_2024,zhang_Floquet_observation_2024,nagulu_chip-scale_2022,zhang_topolectrical_2025}.
	
	\item \textit{Convenience in combining linear and non-linear components}: Electrical circuits provide a diverse range of linear components (such as resistors, capacitors, and inductors) and non-linear components (such as diodes, transistors, and non-linear amplifiers) that can be combined seamlessly. This versatility enables the study and realization of both linear and non-linear phenomena within the same platform~\cite{hadad_self-induced_2018,wang_topologically_2019}. In contrast, photonic and acoustical systems often require specialized materials to achieve non-linearity, making such juxtapositions more challenging~\cite{christodoulides_discretizing_2003,zhang_second_2023}.
\end{itemize}

Although some of these advantages may be even more prominent in certain platforms, i.e., strong intrinsic optical non-linearity in non-linear photonics~\cite{lapine_colloquium_2014,maczewsky_nonlinearity-induced_2020,szameit_discrete_2024,mukherjee_observation_2020,jurgensen_quantized_2021}, electrical circuits remain a highly suitable medium for realizing a broad range of phenomena, as outlined above. In general, their versatility and practicality make them an excellent choice for many applications.

Many of the features listed above have already been effectively exploited in existing topolectrical circuits experiments~\cite{yang_circuit_2024,chen_engineering_2025}. Building upon these advantages, it can be anticipated that the field is moving towards real-world applications of the phenomena investigated. Early attempts, such as sensing device implementations based on topological non-Hermitian phenomena~\cite{de_carlo_non-hermitian_2022}, hint at the potential for further advancements~\cite{deng_ultrasensitive_2024,chen_ultra-sensitivity_2024}. Additionally, the use of analog multipliers to realize complex couplings or temporal responses~\cite{stegmaier_topological_2024} herald a new era when phenomena far beyond topology are realized with chip technologies that are integrated into TE circuits~\cite{kumar_topological_2022,liu_fully_2022,mirhosseini_superconducting_2018}.

\section{Theoretical framework for circuit analysis and impedance properties}

We present a concise review of the fundamentals of TE circuits, focusing on the Laplacian formalism and their impedance characteristics.

\subsection{Laplacian formalism for circuits}
\label{sec:laplacian}

In electric circuit settings, both the Laplacian and Hamiltonian are frequently used. While both are used to describe aspects of the circuit dynamics, they are fundamentally different objects. The circuit Laplacian captures the graph connectivity structure of the circuit by representing the the flow of current in an electrical circuit through Kirchhoff's law. By contrast, the Hamiltonian is formally defined as the generator of time evolution, as in Schr\"odinger's equation. Below, we examine how the Laplacian and Hamiltonian are related, particularly in the context of topolectrical circuit arrays~\cite{lee_topolectrical_2018} where both acquires a momentum-space band description.

For any electrical circuit, the steady-state circuit Laplacian is given by
\begin{equation}
	I = J V,
	\label{EqLaplacian}
\end{equation}
where $V$ represents the vector of node voltages and $I$ denotes the input current vector. The components of each vector are the voltages and current at each node. For instance, in a capacitive topological Su-Schrieffer-Heeger (SSH)~\cite{su_solitons_1979} circuit shown in Fig.~\ref{Fig_Impedance}a, the circuit Laplacian is expressed as
\begin{equation}
	\begin{aligned}
		&	  w_1 V_{n,B} + w_2 V_{n-1,B} = \beta V_{n,A} ,\\
		&	  w_1 V_{n,A} + w_2 V_{n+1,A} = \beta V_{n,B} ,
	\end{aligned}
\end{equation}
where $A$ and $B$ are the sub-lattice nodes in a unit cell $n$; $w_{1}, w_{2}$ represent the admittances of the circuit node connections, and $\beta = w_1 + w_2 + w_g$, with $w_g$ denoting the admittance of the grounding components. For example, in the SSH circuit with inter-cell and intra-cell capacitances $C_1$ and $C_2$, and grounding inductance $L$, the admittances are $w_1 = i\omega C_1$, $w_2 = i\omega C_2$, and $w_g = 1/(i\omega L)$, where $\omega = 2\pi f$ and $f$ is the frequency of the driving signal. While these coupled equations also occur in the commonly known SSH model Hamiltonian, capturing the hoppings between sub-lattice nodes $A$ and $B$ within each unit cell $n$, we emphasize its distinction to the Hamiltonian describing electrical circuits, as described below.

In general, the equation of motion in a circuit setting can be expressed as
\begin{equation}
	\dot{\mathbf{I}}(t) = C \ddot{\mathbf{V}}(t) + U \dot{\mathbf{V}}(t) + L \mathbf{V}(t)  ,
	\label{EqMotions-SecondOrder}
\end{equation}
where the dot and double dot denote the first and second derivatives with respect to time $t$, respectively. The matrices $L$, $U$, and $C$ encode the inductance, conductance, and capacitance from the circuit components, respectively, and collectively define the grounded circuit Laplacian as $J = U + i\omega C + \frac{1}{i\omega} L$. At the resonant frequency, where nonzero voltage signals $\mathbf{V}$ can be sustained without input current, this circuit Laplacian defines the Hamiltonian. With no external input current ($\mathbf{I}(t) = 0$), the circuit dynamics are determined by the homogeneous solutions to Eq.~\eqref{EqMotions-SecondOrder}. To reduce this second-order equation to a first-order system, we define independent variables $\psi_1(t) = \dot{\mathbf{V}}(t)$ and $\psi_2(t) = \mathbf{V}(t)$ and construct the state vector accordingly:
\begin{equation}
	\psi(t) = 
		\begin{pmatrix}
		\psi_1(t) \\ \psi_2(t)
	\end{pmatrix} = 
	\begin{pmatrix}
		\dot{\mathbf{V}}(t) \\ \mathbf{V}(t)
	\end{pmatrix} ,
\end{equation}
and rewrite Eq.~\eqref{EqMotions-SecondOrder} as
\begin{equation}
	C \dot{\psi_1}(t) + U \psi_1(t) +  L \psi_2(t) = 0.
	\label{EqMotions-FirstOrder}
\end{equation}
We can now describe the system with a first-order differential equation for $\psi(t)$:
\begin{equation}
\dv{}{t}	\begin{pmatrix}
		\psi_1(t) \\ \psi_2(t) 
	\end{pmatrix} = 
	\begin{pmatrix}
		-C^{-1} U & -C^{-1} L \\ \mathbb{I} & 0
	\end{pmatrix}
	\begin{pmatrix}
	\psi_1(t) \\ \psi_2(t) 
	\end{pmatrix},
\end{equation}
where $\mathbb{I}$ is the $N \times N$ identity matrix, where $N$ denotes the circuit size. Hence the system dynamics have been reformulated as $-i \dv{}{t}\psi(t) = H \psi(t) $ where the Hamiltonian is just the block matrix
\begin{equation}
	H = i \begin{pmatrix}
		C^{-1} U & C^{-1} L \\ -\mathbb{I} & 0
	\end{pmatrix}.
	\label{EqMotionH}
\end{equation}
This $H$, which is a $2N \times 2N$ block matrix, captures the dynamics of the circuit. The eigenvalues of $H$ correspond to the system’s resonant frequencies, while its eigenvectors describe the (equivalant) spatial and temporal voltage distributions at resonance. $H$ governs the temporal evolution of the circuit represented by the Laplacian $J$. Thus, the steady-state behavior and the time-dependent dynamics of the circuit can be analyzed using Eq.~\eqref{EqLaplacian} and Eq.~\eqref{EqMotionH}, respectively.

\subsection{Impedance in topolectrical circuit arrays}

\begin{figure*}[t!]
	\centering
	\includegraphics[width=\textwidth]{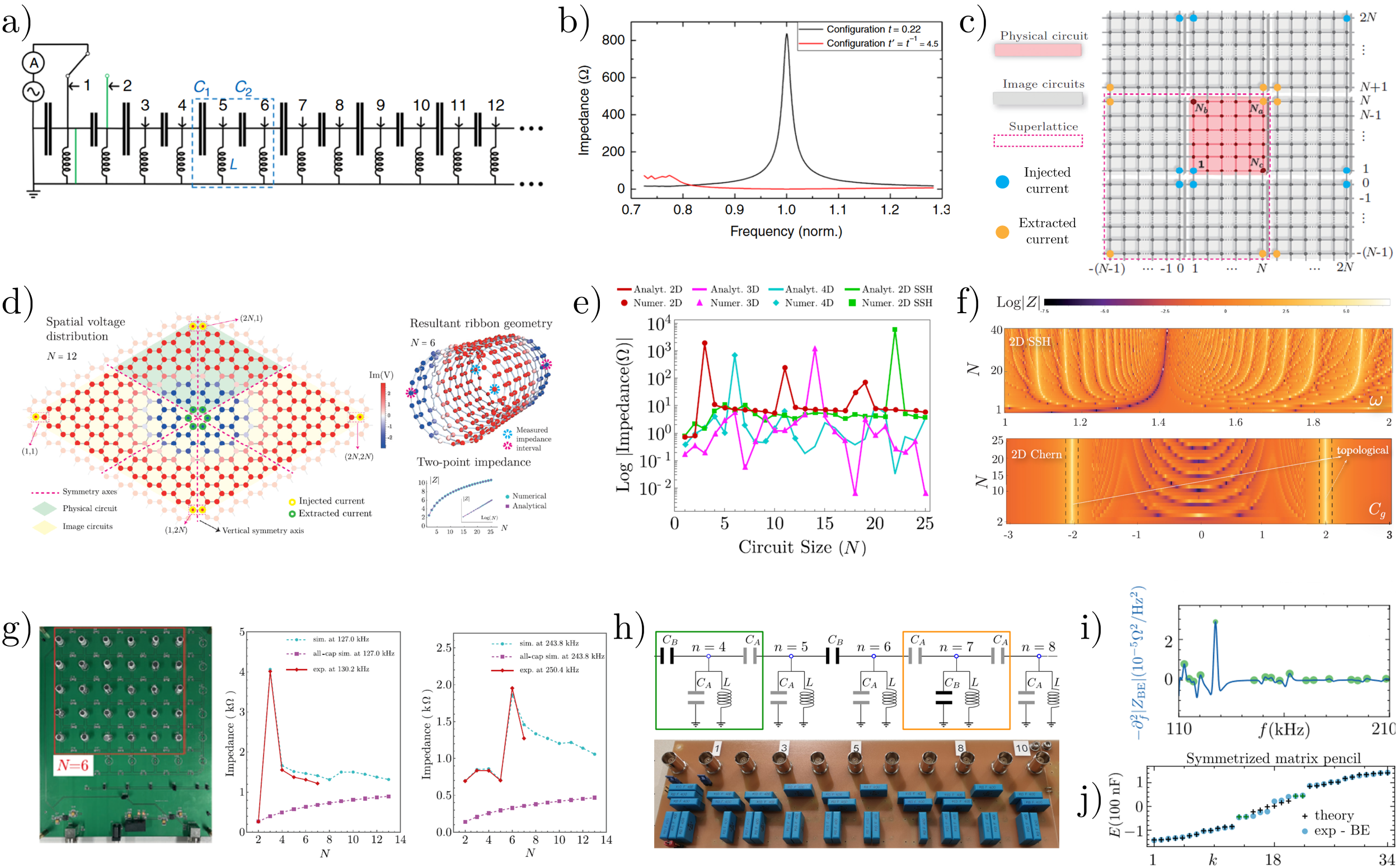}
	\caption{\textbf{Impedance resonances in topolectrical circuits.}  
		\textbf{(a)} The SSH topolectrical circuit consists of capacitors of alternating sizes and on-site inductors, forming a one-dimensional chain with two sites per unit cell~\cite{lee_topolectrical_2018}. 
		\textbf{(b)} Dramatic impedance peaks of the SSH circuit in (a) are observed only in the topologically nontrivial ($t<1$) case, not the trivial ($t>1$) case~\cite{lee_topolectrical_2018}. [(a,b) adapted with permission from Lee et al., Commun Phys 1, 39 (2018). Copyright 2018 licensed under CC BY 4.0.]
		\textbf{(c)} Edge-to-edge or corner-to-corner impedance across a bounded circuit array can be analytically determined using the method of images, which involves infinite periodic tiling of the physical circuit (red shaded) and symmetrically placed current injection (cyan dots) and extraction (orange dots) that makes the physical circuit boundaries equipotentials~\cite{sahin_impedance_2023,mamode_electrical_2017,mamode_laplacian_2024}.
		\textbf{(d)} The method of images can be applied to more generic lattices and geometries too, such as the honeycomb lattice with ribbon geometry~\cite{sahin_impedance_2023}.   
		\textbf{(e)} In heterogeneous circuits with more than one type of elements that give rise to opposite phases, i.e., capacitors and inductors, dramatic impedance peaks can occur at special system sizes where one term in the denominator of Eq.~\eqref{general_Z} nearly vanishes~\cite{zhang_anomalous_2023}.  
		\textbf{(f)} As revealed in Ref.~\cite{zhang_anomalous_2023}, the origin of these abrupt size-dependent resonances can be traced to the fractal-like patterns of the impedance density plot in the parameter space of system size and one of the circuit parameters, such as the AC frequency $\omega$. Resonances arising from the topological zero modes can also appear, although they generally do not scale with system size, as illustrated in the bottom diagram. [(c,d,e,f) adapted with permission from Sahin et al., Phys. Rev. B 107, 245114 (2023). Copyright 2023 by the American Physical Society.]
		\textbf{(g)} The experiment in Ref.~\cite{zhang_anomalous_2023} demonstrates the robustness of these size-dependent peaks against parasitic resistance in a two-dimensional LC circuit. [Adapted with permission from Zhang et al., Commun Phys 6, 151 (2023). Copyright 2023 licensed under CC BY 4.0.]
		\textbf{(h)} The one-dimensional chiral symmetric Fibonacci topolectrical circuit, exhibiting a fractal energy spectrum in the thermodynamic limit, is built to demonstrate how the circuit Laplacian spectrum can be obtained directly from impedance measurements~\cite{franca_impedance_2024}.  
		\textbf{(i)} The peak frequencies (green dots) of the measured impedance between a bulk node and an edge node (blue line) in Ref.~\cite{franca_impedance_2024} can be accurately identified by applying noise filters, such as the Butterworth filter and the second derivative method.  
		\textbf{(j)} As demonstrated in Ref.~\cite{franca_impedance_2024}, the circuit spectrum $E$ can be directly recovered from a single frequency-resolved impedance measurement through the use of signal processing techniques such as the matrix pencil method. [(h,i,j) adapted with permission from Zhang et al., Commun Phys 6, 151 (2023). Copyright 2023 licensed under CC BY 4.0.]
		}
	\label{Fig_Impedance}
\end{figure*}

The impedance is a fundamental measurable quantity that relates the circuit's response to a given excitation. The two-point impedance~\cite{atkinson_infinite_1999,tzeng_theory_2006,essam_exact_2009} between nodes $i$ and $j$ is defined as
\begin{equation}
	Z_{i,j}  = \frac{V_i - V_j}{I},
	\label{EqImpedance}
\end{equation}
which can also be expressed in terms of the eigenvalues and eigenvectors of the Laplacian as~\cite{lee_topolectrical_2018, cserti_application_2000,wu_theory_2004,cernanova_nonsymmetric_2014,izmailian_generalised_2014}
\begin{equation}
	Z_{ij}=\sum_{k, \lambda_k \neq 0} \frac{|\psi_{k i}-\psi_{k j}|^2}{\lambda_k},
	\label{EqImpEig}
\end{equation}
where $\lambda_k$ and $\psi_k$ are the $k$-th eigenvalue and eigenvector of the Laplacian, respectively, and $|...|$ denotes the biorthogonal norm. In the presence of a zero mode, which can exist due to symmetry or topological protection, the impedance would peak. For instance, for the topolectrical circuit realization of the SSH model~\cite{lee_topolectrical_2018} [Fig.~\ref{Fig_Impedance}a], the impedance between two edges exhibits a peak in the topological regime where there exists a $\lambda_k\approx 0$ [black in Fig.~\ref{Fig_Impedance}b], but no resonance peak is observed in the trivial phase [red in Fig.~\ref{Fig_Impedance}b]. While the two-point impedance typically needs to be computed numerically through Eqs.~\eqref{EqImpedance} and \eqref{EqImpEig}, analytical results exists for special scenarios which can be handled through lattice Green's functions~\cite{cserti_application_2000,malik_lattice_2023}, recursion-transform methods~\cite{tan_two-point_2016}, transfer matrix methods~\cite{perrier_two-point_2024}, or, for sufficiently symmetric boundary terminations, the method of images~\cite{sahin_impedance_2023}.

\subsubsection{Method of images for bounded circuit arrays}

A significant obstacle in obtaining impedance results analytically lies in the implementation of open boundary conditions (OBCs), particularly for unit cells with multiple nodes. However, recent developments have demonstrated that the method of images can successfully yield analytical expressions for a wide range of OBC scenarios, for instance 2D SSH circuits with four nodes per unit cell. The core idea involves tiling the space periodically with infinite copies of the actual lattice, such that the open boundary becomes an equipotential where no current flows through anyway. As illustrated in Fig.~\ref{Fig_Impedance}c, symmetrically injecting current at the center nodes and extracting it from the intended boundaries converts the edges of the pink region into impenetrable equipotentials, such that the OBC problem can be treated as a infinite image current problem.

Care has to be taken to design a symmetric manner for current injection and extraction, such that neighboring nodes along the intended boundaries have the same voltage potential. This prevents current from leaking through the physical circuit’s boundaries. This method of images can be performed to induce open boundaries in a variety of lattice geometries, such as the honeycomb lattice [Fig.~\ref{Fig_Impedance}d]~\cite{sahin_impedance_2023}. In the same work, a general analytical expression was derived for the corner-to-corner impedance of a $D$-dimensional circuit with $N^D$ unit cells. For a single node per unit cell, the size $N$-dependent impedance $Z(N)$ is given by:
\begin{equation}
	Z(N)=\frac{2}{N^D} \sideset{}{^*} \sum_{\mathbf{k}} \frac{\left( \prod_{i=1}^D \cos(k_i /2)\right) \times \cos \left(\sum_{i=1}^D k_i /2 \right) }{\left(\sum_{i=1}^D w_{i}(1-\cos(k_i)\right) + w_{gnd}/2},
	\label{general_Z}
\end{equation}
where $w_i$ is the admittance of the nearest-neighbor unit cell coupling along each direction, and  $\mathbf{k}=\sum_{i=1}^{D}k_i \mathbf{a}_i$, with $k_{i}=\frac{n_i\pi}{N}$ and $n_i\in\{1,2,...,2N\}$, where $i=(1,2,\dots,D)$. Here the sum is taken for $n_i$ up to $2N$, not $N$, because of the doubling of the system to account for the image nodes. $w_{gnd}$ is the admittance of the uniform grounding connection from each node, introduced such as to provide an offset in the denominator. The asterisk sign $(^*)$ on the summation operator indicates that the impedance computation must be performed for only odd values of $(n_1+n_2+\dots+n_D)$.

Importantly, $Z(N)$ can be extraordinarily large for certain some values of certain circuit sizes $N$ if there are circuit coupling elements that introduce opposite phases, i.e., capacitors and inductors [Fig.~\ref{Fig_Impedance}e]. This is because having at least two different $w_i$ of opposite signs can allow the denominator to almost vanish at appropriate values of $N$. Such dramatic system size dependence in the impedance is not seen in homogeneous electrical media, where the dependence on $N$ is smooth (logarithmic or power-law)~\cite{cserti_application_2000,mamode_revisiting_2021,asad_infinite_2013,mamode_calculation_2019}.

Behind the seemingly erratic occurrence of impedance peaks in Fig.~\ref{Fig_Impedance}e is a sophisticated fractal-like pattern in the distribution of resonance heights within the parameter space of system size $N$ and $\omega$---the dimensionless (normalized) frequency characterized by the LC components [Fig.~\ref{Fig_Impedance}f]. While the detailed pattern depends on the actual lattice model describing the circuit array, hierarchies of impedance peaks universally occur in a qualitatively similar manner. The intricacy of this pattern arises from the commensurability properties of the denominator in Eq.~\eqref{general_Z}, in which some values of $N$ can cause it to be especially close to vanishing [Fig.~\ref{Fig_Impedance}g]. Notably, there exists some branches that are particularly robust and hence experimentally detectable~\cite{zhang_anomalous_2023}, reminiscent of the more salient structures in the Hofstadter butterfly. Additionally, resonances may also arise due to protected topological zero modes, as in a Chern lattice [bottom diagram in Fig.~\ref{Fig_Impedance}f].

\subsection{Extracting the Laplacian spectrum from impedance resonances}
The Laplacian eigenspectrum can be obtained either by reconstructing the Laplacian matrix and then diagonalizing it, or by directly measuring the impedance resonances. 

\subsubsection{Through circuit Laplacian reconstruction}

In the former, one can perform $N^2$ impedance measurements by separately subjecting the circuit to $N$ different input current configurations and measuring the electrical potentials at all the $N$ nodes~\cite{zou_observation_2021,zhang_observation_2024_2,shang_observation_2024,guo_scale-tailored_2024,liu_non-hermitian_2021}. This yields $N$ different current and voltage vectors, which can be written as $N\times N$ matrices $\mathbf{I}$, $\mathbf{V}$. 
This allows the Laplacian $J$ to be recovered via $J = \mathbf{I V^{-1}}$; for simplicity, one can just ground the circuit and choose to input the current at one node at a time, such that $\mathbf I$ is simply the identity matrix.

This is demonstrated in Helbig et al.~\cite{helbig_generalized_2020}, where the eigenvalue band structure of a 1D non-Hermitian SSH circuit under both open and periodic boundary conditions (OBCs and PBCs) was measured. By varying the injected signal frequency and measuring voltage responses, they extracted the system’s eigenvalue spectrum as a function of the driving frequency.
For PBCs, translational symmetry reduces the independent degrees of freedom from $N^2$ to $N$, since it allows the spectrum to be computed in momentum space by exciting the nodes and Fourier transforming the measured node voltages~\cite{stegmaier_topological_2021,li_ideal_2021}. However, OBC circuits require real-space voltage measurements across the entire array, yielding for instance additional spatially localized edge modes. Helbig et al.~\cite{helbig_band_2019} first presented this framework, with a further implementation in a 2D topological Chern circuit by Hofmann et al.~\cite{hofmann_chiral_2019}.

\subsubsection{Through direct circuit resonance measurement}

Constructing the circuit Laplacian in full requires $N^2$ impedance measurements in general, and that presents practical limitations in larger circuits. Recently, Franca et al.~\cite{franca_impedance_2024} developed a method to extract the Laplacian eigenenergies directly from impedance resonances using much fewer measurement configurations. As deduced from Eq.~\eqref{EqImpEig}, any vanishing eigenvalue corresponds to a large impedance resonance. By tuning the system across a range of AC frequencies, various resonances can be observed, as demonstrated with high accuracy in their Fibonacci circuit implementation [Fig.~\ref{Fig_Impedance}h]. This approach bypasses the need for constructing the circuit Laplacian, reducing the measurement complexity from $N^2$ measurements to a single impedance measurement at various (easily tunable) frequencies.

To elaborate, this approach involves examining the second derivative of the impedance profile with respect to frequency, coupled with a fourth-order Butterworth filter to mitigate noise [Fig.~\ref{Fig_Impedance}i]. Advanced data processing techniques, such as the matrix pencil method and utilizing the symmetric energy profiles of chiral symmetric systems, further enhance the precision in determining admittance eigenvalues. In a similar spirit as other studies that utilize machine learning to infer the spectrum from limited measurements (to be discussed in subsequent sections, e.g., Ref.~\cite{shang_experimental_2022}), this method represents a significant advance which allows for the the measurement of the Laplacian spectrum from a single measurement [Fig.~\ref{Fig_Impedance}j].

\section{Circuit implementations of non-Hermitian phenomena}

In the following, we highlight various works, mostly experimental, on demonstrating non-Hermitian phenomena of contemporary interest. Emphasis is placed on recent developments in the last few years.

\subsection{Exceptional point physics}
	
\subsubsection{Exceptional points in parity-time symmetric setups}

\begin{figure*}[ht!]
	\centering
	\includegraphics[width=\textwidth]{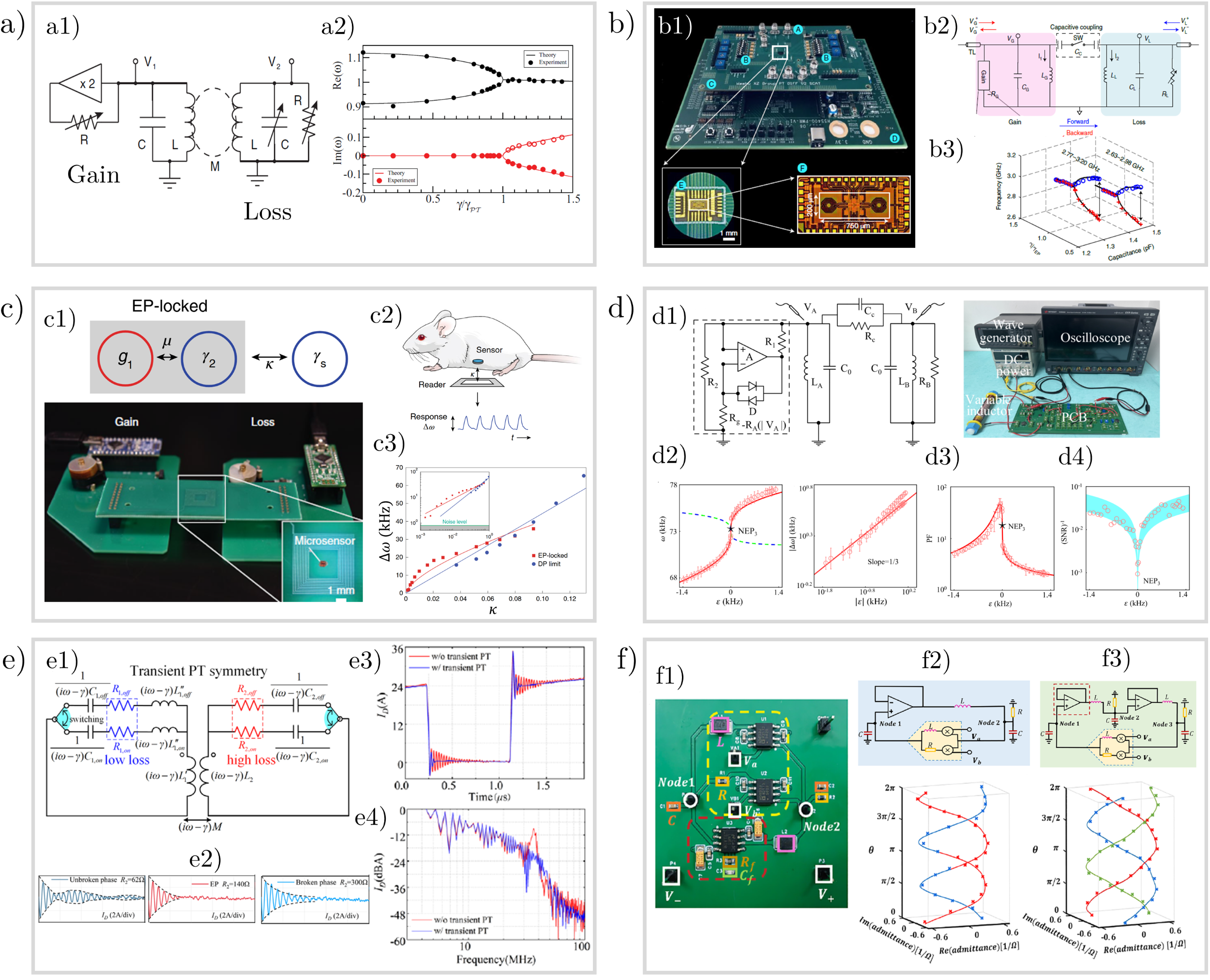}
	\caption{\textbf{Parity-time (PT) symmetric two-component systems and their circuit realizations.}  
		\textbf{(a)} An early demonstration of a parity-time symmetric RLC circuit (a1) exhibiting a complex branch point, i.e., EP (a2) in the frequency spectrum, as the mutual coupling between two inductors is tuned~\cite{schindler_experimental_2011}. [Adapted with permission from Schindler et al., Phys. Rev. A 84, 040101(R) (2011). Copyright 2011 by the American Physical Society.]
		\textbf{(b)} An integrated circuit realization of a PT-symmetric two-component system that demonstrates sensitivity to coupling capacitance, as seen in the 0.35 GHz band isolation corresponding to peak splitting in the frequency domain (b3)~\cite{cao_fully_2022}. [Adapted with permission from Cao et al., Nat. Nanotechnol. 17, 262–268 (2022). Copyright 2022 licensed under CC BY 4.0.]
		\textbf{(c)} Implementation of such two-component systems as microsensors, with wireless systems locked to an exceptional point (EP). The EP-locked setting (red in c3) shows superior sensitivity compared to the conventional diabolic point response (DP limit shown in blue in c3)~\cite{dong_sensitive_2019}. [Adapted with permission from Dong et al., Nat Electron 2, 335–342 (2019). Copyright 2019 by Springer Nature.]
		\textbf{(d)} Enhanced sensitivity at the EP from introducing non-linearity. Such non-linear EPs (NEPs) (d2) enjoy greater noise resilience (d3 and d4) due to the restoration of a complete eigenstate due to the non-linearity~\cite{bai_observation_2024}. [Adapted with permission from Bai et al., Phys. Rev. Lett. 132, 073802 (2024). Copyright 2024 by the American Physical Society.]
		\textbf{(e)} PT-symmetric systems can enhance stability in the transient domain beyond their sensitive response. In the presence of transient PT symmetry, unwanted oscillations (red in e3) can be eliminated in fast-switching devices. The vanishing peak around 36 MHz in e4 indicates the suppression of unwanted oscillations~\cite{yang_observation_2022}. [Adapted with permission from Yang et al., Phys. Rev. Lett. 128, 065701 (2022). Copyright 2022 by the American Physical Society.]
		\textbf{(f)} Higher-order EPs underlie braiding in the complex eigenvalue plane. The spectra of the two-node (f2) and three-node (f3) circuits exhibit two-band and three-band twistings hosting three EPs~\cite{cao_observation_2024}. [Adapted with permission from Cao et al., Phys. Rev. B 109, 165128 (2024). Copyright 2024 by the American Physical Society.] }
	\label{Fig_PT}
\end{figure*}

Parity-time (PT) symmetry can protect the reality of eigenvalues even if the operator, i.e., circuit Laplacian or Hamiltonian is non-Hermitian~\cite{el-ganainy_non-hermitian_2018,wiersig_enhancing_2014,zhao_exceptional_2024,chen_generalized_2018,chen_exceptional_2017,choi_observation_2018}. A typical PT-symmetric 2-level Hamiltonian is represented as
\begin{equation}
	H_{\mathrm{PT}} = 
	\begin{pmatrix}
		\omega + i\gamma & \kappa \\
		\kappa & \omega - i\gamma
	\end{pmatrix},
\end{equation}
where $\pm \gamma$ are the gain and loss terms, $\omega$ is the natural frequency of both subsystems, and $\kappa$ is their coupling strength. In linear electrical circuits, the most common operator of interest would be the circuit Laplacian instead of the Hamiltonian.

The eigenvalues of $H_{\mathrm{PT}}$ are given by $E = \omega \pm \sqrt{\kappa^2 - \gamma^2}$. When $\kappa > \gamma$, the eigenvalues are real, indicating the PT-symmetric phase. Conversely, when $\kappa < \gamma$, the eigenvalues become complex, signaling a transition to the broken PT-symmetric phase. At the critical point $\kappa = \gamma$, not only do the eigenvalues coincide, but the eigenvectors also coalesce (become parallel to each other). Such non-Hermitian critical points are known as exceptional points (EPs)~\cite{wiersig_sensors_2016,wiersig_review_2020,ding_non-hermitian_2022,ozdemir_paritytime_2019,meng_exceptional_2024,miri_exceptional_2019,hodaei_enhanced_2017,chen_exceptional_2017}, which are special because the eigenspace is defective, i.e., not full rank. Further, due to the square-root singularity of the eigenbands at the EP, the spectrum becomes highly sensitive to perturbations, with a divergent gradient as the EP is approached. This sensitivity increases for higher-order exceptional points, where $n>2$ eigenvalues coincide and result in a $E^{1/n}$ singularity. That the system's response to perturbations becomes increasingly pronounced is particularly advantageous for sensing applications~\cite{hodaei_parity-timesymmetric_2014,peng_paritytime-symmetric_2014}. The sensitivity can be enhanced by incorporating multiple modes or alternative coupling mechanisms to enable higher-order EPs~\cite{hodaei_enhanced_2017,lee_topologically_ep_2024,gohsrich_exceptional_2024,panahi_higher-order_2024,xiao_experimental_2023,li_stochastic_2023,hu_unconventional_2025}. Beyond isolated EPs, higher dimensional extensions such as exceptional lines~\cite{tang_realization_2023} and exceptional rings~\cite{cao_band_2020} have also been observed, in analogy to nodal lines and rings~\cite{burkov_topological_2011,hu_transport_2019,weng_topological_2015,bian_drumhead_2016,fang_topological_2015,wang_experimental_2022}.

One of the earlier experimental realizations of a PT-symmetric circuit was performed by Schindler et al.~\cite{schindler_experimental_2011}, where two LC oscillators---one with gain (achieved using an operational amplifier) and the other with loss (introduced by a resistor)---were coupled [Fig.~\ref{Fig_PT}a1]. These systems demonstrate the square-root branching behavior that leads to the bifurcation of both the real and imaginary parts of the eigenfrequency at the EPs [Fig.~\ref{Fig_PT}a2].

Integrated circuit implementations of PT-symmetric systems are particularly technologically promising, as they operate at high microwave frequencies and offer scalability. For instance, Cao et al.~\cite{cao_fully_2022} designed a fully integrated PT-symmetric two-component system [Fig.~\ref{Fig_PT}b1] where the oscillatory mode exhibits wide-band microwave generation with significantly reduced noise aided by the PT-symmetric phase transition [Fig.~\ref{Fig_PT}b2]. The splitting of resonance peaks in different phases achieved a 0.35 GHz band isolation [Fig.~\ref{Fig_PT}b3], showcasing the exceptional sensitivity and practical utility of integrated circuit-based PT-symmetric systems.

Wireless electronic sensors~\cite{chen_inductor-capacitor_2024} primarily utilize the resonant behavior of LC circuits [Fig.~\ref{Fig_PT}c1], and enhancing their sensitivity can lead to promising applications such as wireless power transfer~\cite{assawaworrarit_robust_2020}. Also, implanted LC wireless circuits can monitor biological functions in living organisms [Fig.~\ref{Fig_PT}c2]. By leveraging on the highly sensitive readout of their microsensor implanted in vivo, Dong et al.~\cite{dong_sensitive_2019} successfully measured physiological functions such as breathing rate, demonstrating superior sensitivity compared to standard LC sensory readers [Fig.~\ref{Fig_PT}c3].

Although an EP system becomes highly responsive at the EP, the accompanying defective eigenspace (where two or more eigenvectors coalesce) can also introduce undesirable susceptibility to noise~\cite{bai_nonlinear_2023}, complicating the implementation of EP-based sensors~\cite{wiersig_review_2020}. However, appropriately designed non-linearity can mitigate that susceptibility while preserving the responsiveness, even leading to EP transitions~\cite{dai2024non,gong2024topological}. Bai et al.~\cite{bai_observation_2024} designed a two-component circuit that juxtaposes the non-linearity and non-Hermitian coupling between two oscillators [Fig.~\ref{Fig_PT}d1]. Introducing non-linear gain results in higher-order non-linear EPs (NEPs), such as third-order NEPs, as shown in Fig.~\ref{Fig_PT}d2. They effectively suppress noise, even when the circuit is intentionally disturbed by a non-negligible external pulse~\cite{fang2024exceptional}. With non-linearity, the Petermann factor can be prevented from diverging [Fig.~\ref{Fig_PT}d3]. Additionally, the signal-to-noise ratio reaches its lowest value at the NEP and increases as the system moves away from the NEP [Fig.~\ref{Fig_PT}d4].

PT symmetry-based implementations can also enhance system stability. In semiconductor technologies, fast-switching devices are widely used, and achieving both high speed and stability is highly desirable~\cite{chen_review_2020,rodrigues_review_2021}. Yang et al.~\cite{yang_observation_2022} demonstrated that transient PT symmetry, as implemented in their electric circuit [Fig.~\ref{Fig_PT}e1], can significantly reduce unwanted oscillations during fast switching. Noise-induced perturbations in the oscillations decay more rapidly at the EP compared to in the broken or unbroken PT phases [Fig.~\ref{Fig_PT}e2]. For step-like dynamic switching, transient PT symmetry effectively suppresses unwanted oscillations [red in Fig.~\ref{Fig_PT}e3], as observed from the diminishing peak around 36 MHz in the transient PT phase [blue in Fig.~\ref{Fig_PT}e4] in the frequency domain.

Mathematically, EPs can also trace out interesting mathematical structures in either parameter space or real/momentum space. EPs underlie braiding in the complex eigenvalue plane~\cite{zhu_versatile_2024,pu_non-hermitian_2025} and define knots in the band topology~\cite{bi_nodal-knot_2017}. By designing non-Hermitian circuits involving multiple bands, Cao et al.~\cite{cao_observation_2024} achieved higher-order EPs. The shape of knots or braids primarily depends on the winding paths around the EPs~\cite{ryu_exceptional_2024}. For example, a circuit with two nodes [Figs.~\ref{Fig_PT}f1 and f2] or three nodes [Fig.~\ref{Fig_PT}f3] can exhibit two-component or three-component knot topology, with double or triple twisting in the complex eigenenergy plane. Momentum-space EP knots are also described in Refs.~\cite{zhang_tidal_2021,carlstrom_knotted_2019,stalhammar_hyperbolic_2019}. In a larger parameter space, multi-band EPs can also describe mathematical singularities known as catastrophes, which have deep connections, i.e., McKay correspondence with diverse mathematical constructs such as the ADE classification of Lie algebras, finite subgroups of $SU(2)$, platonic solids and the monodromy group of simple singularities~\cite{arnold1986catastrophe,chandrasekaran_catastrophe_2020}. A recent circuit experiment has mapped out the swallowtail catastrophe in parameter space~\cite{hu_non-hermitian_2023}.

\subsubsection{Exceptional bound states as robust circuit resonances}

\begin{figure}[ht!]
	\centering
	\includegraphics[width=\linewidth]{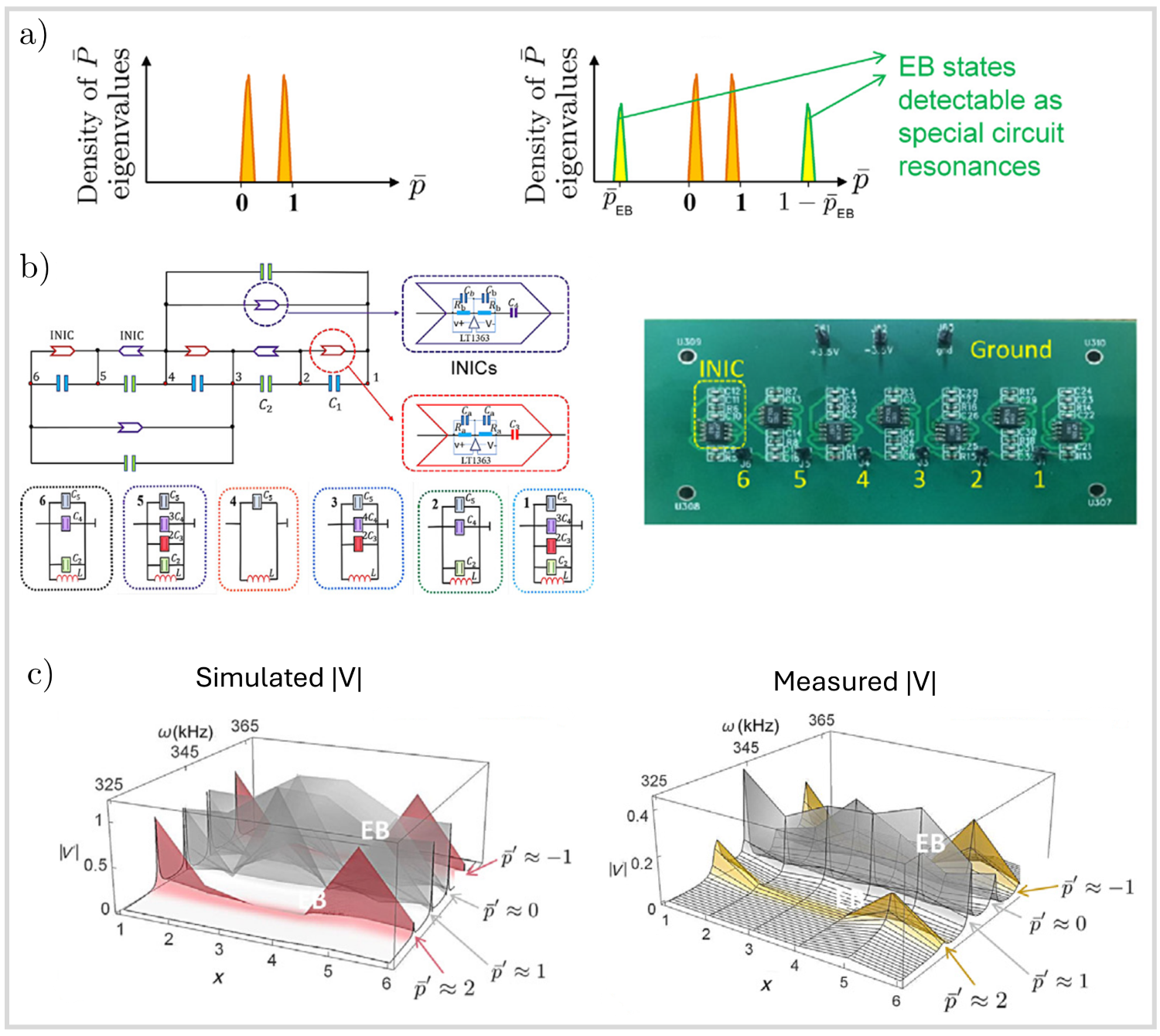}
	\caption{\textbf{Observation of exceptional bound (EB) states in electrical circuits.} 
		\textbf{(a)} Exceptional bound states in the spectrum $\bar p$ of the truncated two-point function matrix arise robustly from eigenspace defectiveness at EPs. When implemented in an electrical circuit, they show up as strong isolated circuit resonances (yellow) in addition to the usual peaks near $0$ and $1$.
		\textbf{(b)} Schematic of the circuit in implementing the circuit Laplacian hosting EB states, as studied in Ref.~\cite{zou_experimental_2024}. op-amps are used to realize the asymmetric couplings.
		\textbf{(c)} Simulated and measured node voltages as a function of AC frequency $\omega$ exhibit voltage profiles corresponding to the spatial profile of the EB states at resonant frequencies corresponding to the two EB eigenvalues, $\bar{p}^{'} \approx -1$ and $\bar{p}^{'} \approx 2$.
		[(a,b,c) adapted with permission from Zou et al., Science Bulletin 69, 2194 (2024). Copyright 2024 Science China Press.]}
	\label{Fig_EBstates}
\end{figure}

The defectiveness of EPs causes the occupied band projector to be singular, leading to substantial non-locality in its two-point correlation functions. A prominent consequence is the appearance of special isolated eigenstates [Fig.~\ref{Fig_EBstates}a] of the real-space truncated two-point correlator matrix known as exceptional bound (EB) states~\cite{lee_exceptional_2022,zou_experimental_2024,liu_non-hermitian_2024, xue_topologically_2024}. These EB states differ from topological modes, which are protected by non-trivial bulk topology, and non-Hermitian skin modes, which stem from non-trivial point gap band topology and are also boundary-localized in open-boundary systems~\cite{gong_topological_2018,li_geometric_2019,kawabata_symmetry_2019,ashida_non-hermitian_2020,lin_topological_2023,okuma_non-hermitian_2023,zhang_review_2022,yokomizo_non-bloch_2022}. While most eigenvalues $\bar p$ of the truncated two-point correlator matrix remain clustered around $0$ or $1$, as in non-EB cases, a distinctive pair of EB states $\bar p_\text{EB}$ and $1-\bar p_\text{EB}$ additionally appears as long as the parent Hamiltonian is defective [Fig.~\ref{Fig_EBstates}a]. The value of $\bar p_\text{EB}$ scales strongly with the system size, as detailed in Refs.~\cite{lee_exceptional_2022,zou_experimental_2024}. Since the spectrum of the truncated two-point correlator matrix corresponds to fermionic occupation probabilities, EB states also lead to negative entanglement entropy in the context of free fermions~\cite{lee_exceptional_2022,zou_experimental_2024,lee_free-fermion_2015}, with interesting ramifications in topological~\cite{xue_topologically_2024} or competitive non-Hermitian skin effect settings~\cite{qin_universal_2023}.

Experimentally, EB states can be simulated in classical platforms that mathematically describe the two-point correlator matrix. Electrical circuits are particularly suited for this purpose because any two nodes can be connected at will, free from the constraints of locality. As a pioneering demonstration, the circuit [Fig.~\ref{Fig_EBstates}b] designed by Zou et al.~\cite{zou_experimental_2024} successfully demonstrated the robustness of EB states through measured voltage profiles [Fig.~\ref{Fig_EBstates}c]. Strong resonances at frequencies corresponding to the eigenvalues of the EB states were detected, demonstrating the direct measurability of these states. These EB states are shown to be much more robust than the non-EB resonances, which can be significantly disturbed by relatively minute perturbations of the larger hoppings~\cite{zou_experimental_2024}.

\subsection{Non-Hermitian skin effect}
\label{sec:NHSE}

One of the most intensely studied non-Hermitian phenomenon is the the non-Hermitian skin effect (NHSE), where states accumulate against spatial inhomogeneities, impurities or boundaries~\cite{li2025phase,lee_anomalous_2016,yang_non-hermitian_2020,martinez_alvarez_non-hermitian_2018,yao_edge_2018,lee_anatomy_2019,kawabata_symmetry_2019,ashida_non-hermitian_2020,gong_topological_2018,li_non-hermitian_2022,zhang_observation_2024_2,lieu_topological_2018,jiang_dimensional_2023,li_parity-dependent_2024,he_evanescent_2024,lee_unraveling_2020,qi_extended_2024,rafi-ul-islam_dynamic_2024,yuce_strong_2024,li_spatially_2024,li_exact_2025,rafi-ul-islam_non-hermitian_2021,oztas_su-schrieffer-heeger_2019,ghatak_observation_2020,yuce_edge_2018,xue_non-hermitian_2022,lee_hybrid_2019,li_gain-loss-induced_2022,kawabata_higher-order_2020,longhi_self-healing_2022,kunst_biorthogonal_2018,fu_degeneracy_2022,zhang_observation_2025,arouca2020unconventional,lu2021magnetic,jiang2022filling,yang2022designing,longhi2022non,zhang2022real,gu2022transient,shen2023proposal,qin2023non,qin2024kinked,li2024observation,wang2024non,gliozzi2024many,yoshida2024non,liu2024localization,zhao2025two,hamanaka2025multifractality,shen2024enhanced,yang2024percolation,shen2025observation,kim_boundary-driven_2025,wang_tunable_2025}, challenging conventional condensed-matter insights into the correspondence between bulk and boundary properties. 

In most cases, the NHSE occurs when the periodic boundary condition (PBC) spectrum encloses a finite area in the 2D complex spectral plane, which generally occurs when asymmetric couplings are present. Then, it can be shown that the open boundary condition (OBC) spectrum must assume the form of arcs or branches within the PBC spectral loop. Referring the reader to the many excellent references that elaborate on this~\cite{borgnia_non-hermitian_2020,okuma_topological_2020,li_geometric_2019,lin_topological_2023,tai_zoology_2023,guo_exact_2021,yokomizo_non-bloch_2022,yao_edge_2018,longhi_probing_2019,jiang_dimensional_2023,li_universal_2024,xiong_graph_2024}, we briefly note that the OBC spectrum, and hence the NHSE, can be understood from the Bloch PBC Hamiltonian $H(k)$ by complex-deforming the momentum, i.e., introducing an imaginary flux~\cite{lee_anatomy_2019,xiong2018does,lee_unraveling_2020,xiong_graph_2024}:
\begin{equation}
	H_\kappa(k) = H(k+i\kappa(k)),
\end{equation}
where $k$ and $\kappa$ represent the real and imaginary components of the momentum wavevector, respectively. The form of the complex deformation $\kappa(k)$, which generically depends on $k$, is crucial for understanding the nature of the so-called skin modes which exponentially localize against the system's boundaries. For any Bloch wavevector $k$ with eigenvector $\psi_k$, an appropriate complex momentum deformation $k\rightarrow k+i\kappa(k)$ would yield the OBC eigenequation
\begin{equation}
H_\kappa(k)\psi_{k+i\kappa(k)}= E_\text{OBC}\psi_{k+i\kappa(k)},
\label{GBZ0}
\end{equation}
where $E_\text{OBC}$ belongs to the OBC spectrum. Here $\psi_{k+i\kappa(k)}(x)\sim e^{-\kappa(k)x}$ is the complex momentum-deformed eigenstate that tends towards the OBC eigenstate in the thermodynamic limit, with $\kappa(k)$ having the physical interpretation of the inverse skin decay length. From Eq.~\eqref{GBZ0}, $E_\text{OBC}$ and its corresponding $\kappa(k)$ are related via the characteristic equation
\begin{equation}
	\det\abs{H(k+i\kappa(k))-E_\text{OBC}}=0.
	\label{GBZ}
\end{equation}
Because $E_\text{OBC}$ is in general not in the Bloch (PBC) spectrum, $\kappa(k)$ is in general nonzero too. To determine values of $E_\text{OBC}$ which are actually allowed to exist in the OBC spectrum, we note that satisfying OBCs at both boundaries simultaneously generally requires the exact OBC eigenstate to be a superposition of two different bulk solutions with equal decay lengths $\kappa^{-1}$. In other words, the OBC eigenenergies are those values of $E_\text{OBC}$ for which Eq.~\eqref{GBZ} simultaneously admits two different $k$ solutions with the same value of $\kappa(k)$, a condition famously known as the GBZ condition~\cite{yang_non-hermitian_2020,wang_amoeba_2024,song_non-hermitian_2019,kawabata_non-bloch_2020,xiong_graph_2024}.

Note that the GBZ condition, as well as the existence of a unique $\kappa(k)$ that quantifies the skin accumulation, may no longer hold when more than one NHSE channel coexists in the system. A prominent example is the critical 
non-Hermitian skin effect (cNHSE)~\cite{li_critical_2020,rafi-ul-islam_critical_2022,siu_terminal-coupling_2023,qin_universal_2023,zhang2024algebraic,yang2024tailoring,liu_non-hermitian_2024,rafi-ul-islam_critical_2025}, where weakly coupled chains with different NHSE directions produces feedback loops that amplify certain states and lead to qualitatively different band structure descriptions at different system sizes. Appropriately designed cNHSE systems may even undergo an exceptional phase transition at a particular system size, with special scaling-induced EP defectiveness and leading to unique entanglement dips~\cite{liu_non-hermitian_2024}. Fundamentally, the GBZ condition hinges on the assumption that the OBC eigenstate only comprise of the superposition of two bulk solutions; relaxing that generally leads to a ``fragmented'' GBZ in which more than one $\kappa(k)$ participates in the physics substantially~\cite{qin_occupation-dependent_2024,wang_non-hermitian_2025}.

\begin{figure*}
	\centering
	\includegraphics[width=17cm]{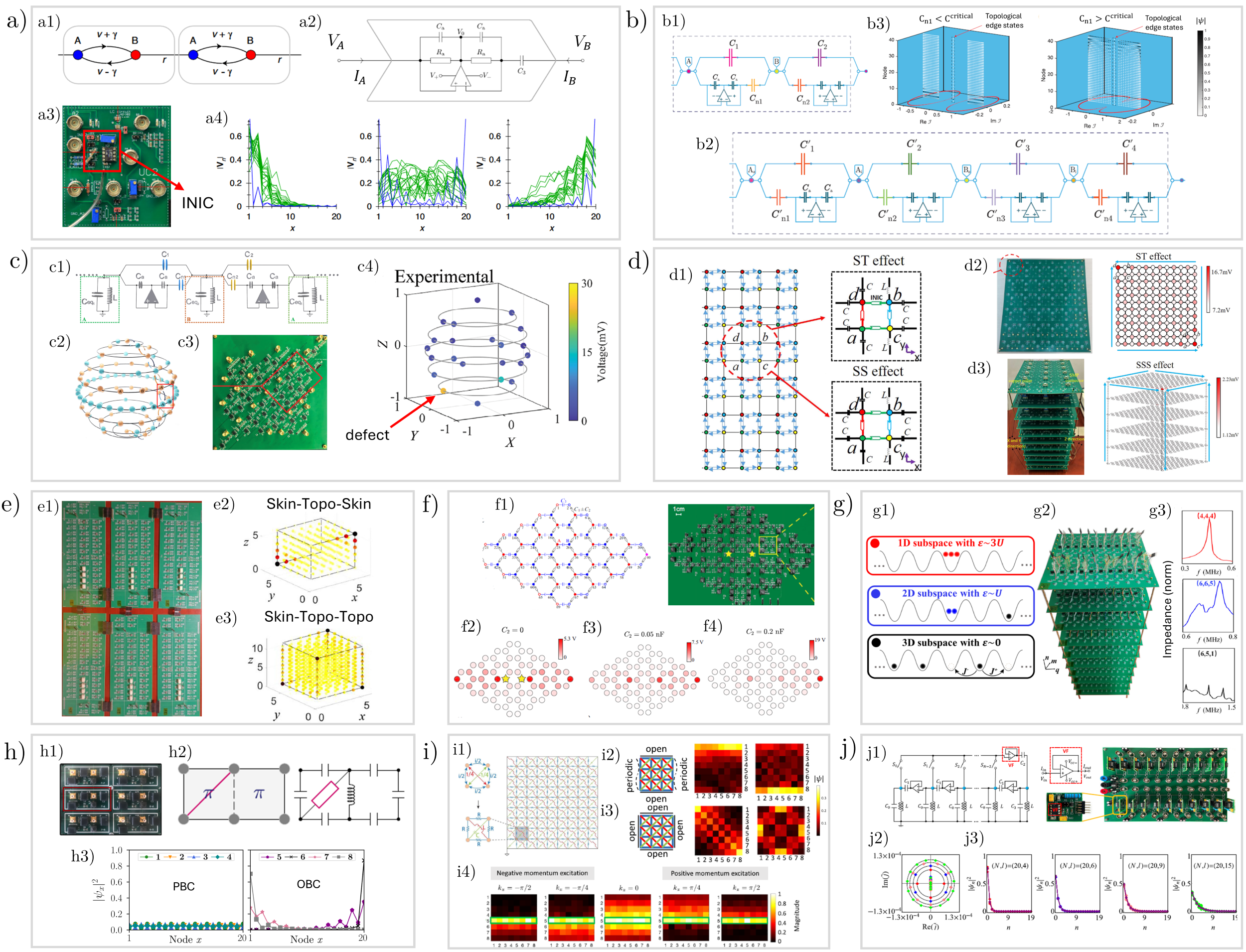}
	\caption{\textbf{Circuit realizations of non-Hermitian skin effect (NHSE) phenomena.}  
		\textbf{(a)} First experimental realization of the NHSE using impedance converters (a2) for non-reciprocal couplings. Voltage is localized either to the left or right when coupling asymmetry is broken (left and right in a4, respectively), while localization is absent when reciprocity is restored (middle in a4)~\cite{helbig_generalized_2020}. [Adapted with permission from Helbig et al., Nat. Phys. 16, 747–750 (2020). Copyright 2020 under CC BY 4.0.]
		\textbf{(b)} Control of the NHSE in a multi-component circuit with multiple asymmetric couplings. Mode localization (darker colors indicate larger eigenmode magnitudes) can still occur in models with multiple asymmetric couplings by setting one of the asymmetric couplings ($C_{\text{n1}}$ in b1 and b3) relative to the critical value~\cite{rafi-ul-islam_unconventional_2022}. [Adapted with permission from Rafi-Ul-Islam et al., Phys. Rev. Research 4, 043108 (2022). Copyright 2022 under CC BY 4.0.]
		\textbf{(c)} Circuit realization of NHSE in a spherical geometry, exhibiting a localized profile around a topological defect. The elevated voltage measured at the defective node, shown in c4, demonstrates defect-induced localization~\cite{shen_energy_2024}. [Adapted with permission from Shen et al., Annalen der Physik, 536(11), 2400026 (2024). Copyright 2024 by Wiley-VCH GmbH.]	
		\textbf{(d)} Hybrid higher-order NHSE in non-reciprocal 2D and 3D topolectrical lattices, where topological localization in one direction can control the skin localization in a transverse direction. The skin-topological (ST effect) and skin-skin-skin (SSS effect) hybrid higher-order effects are observed as localized voltages at the corners shown in d2 and d3, scaling with different co-dimensions~\cite{zou_observation_2021}. [Adapted with permission from Zou et al., Nat Commun 12, 7201 (2021). Copyright 2021 under CC BY 4.0.]	
		\textbf{(e)} Hybrid skin-topological NHSE can be harnessed as topological switches for the NHSE, where the extent of resistive loss is used to drive topological transitions which in turn activate or inactivate the NHSE in a transverse direction. The eigenmode localization of skin-topo-skin and skin-topo-topo cases, shown in e2, exhibit two and four corner localization behaviors~\cite{zhang_electrical_2023}. [Adapted with permission from Zhang et al., Phys. Rev. B 107, 085426 (2023). Copyright 2023 by the American Physical Society.]
		\textbf{(f)} Control of topological modes by the NHSE in a honeycomb lattice. The topological mode is dragged by NHSE as the capacitance parameter $C_2$ increases, as shown in f4~\cite{tang_competition_2023}. [Adapted with permission from Tang et al., Phys. Rev. B 108, 035410 (2023). Copyright 2023 by the American Physical Society.]
		\textbf{(g)} Realization of particle interaction dynamics in a classical circuit array by mapping the 1-dimensional degrees of freedom of multiple particles onto the degrees of freedom of a single particle in multiple dimensions. %Interaction-induced non-Hermitian aggregation effects are observed as impedance resonances at Hilbert space boundaries.
		The impedance peaks shown in g3, while approaching the effective boundaries, demonstrate skin accumulation toward the boundary~\cite{zhang_observation_2022}. [Adapted with permission from Zhang et al., Phys. Rev. B 105, 195131 (2022). Copyright 2022 by the American Physical Society.]		
		\textbf{(h)} Reciprocal skin effect in a 2D fully passive electrical circuit with diagonal loss elements introducing non-Hermiticity. The inverse participation ratio shows no localized states (left in h3) under periodic boundary conditions, but skin states are present under OBCs (right in h3)~\cite{hofmann_reciprocal_2020}. [Adapted with permission from Hofmann et al., Phys. Rev. Research 2, 023265 (2020). Copyright 2020 by the American Physical Society.] 
		\textbf{(i)} NHSE driven by higher-rank chirality in a 2D fully passive electrical circuit, where a 1D chiral mode is most amplified~\cite{zhu_higher_2023}. [Adapted with permission from Zhu et al., Nat Commun 14, 720 (2023). Copyright 2023 under CC BY 4.0.]
		\textbf{(j)} Scale-tailored non-Hermitian localization realized in a circuit with long-range asymmetric couplings~\cite{guo_scale-tailored_2024}. %The spatial profiles of eigenstates show perfect overlap for different combinations of tail and circuit sizes but become less localized as the tail size $l$ increases~\cite{guo_scale-tailored_2024}. 
		[Adapted with permission from Guo et al., Nat Commun 15, 9120 (2024). Copyright 2024 under CC BY 4.0.]}
	\label{Fig_NH}
\end{figure*}

\subsubsection{Circuit demonstrations of one-dimensional NHSE phenomena}
\label{sec:NHSE:multiple}

Extensive eigenmode localization was first observed experimentally in a non-Hermitian electrical circuit implementation. Helbig et al.~\cite{helbig_generalized_2020} constructed an SSH circuit [Fig.~\ref{Fig_NH}a1], where intra-cell couplings were made non-reciprocal using negative impedance converters with current inversion (INICs), as shown in Fig.~\ref{Fig_NH}a2. The INIC changes the phase of the component connected to the inverting (negative) pole of the op-amp by 180 degrees, producing an effective negative impedance from right to left and positive impedance from left to right. The two-node Laplacian of this INIC circuit takes the form
\begin{equation}
	\begin{pmatrix}
		I_{AB} \\ I_{BA}
	\end{pmatrix} = i\omega \gamma 
	\begin{pmatrix}
		-1 & 1 \\ -1 &1 
	\end{pmatrix}
	\begin{pmatrix}
		V_A \\ V_B
	\end{pmatrix},
\end{equation}
where $\gamma$ represents the magnitude of the non-reciprocal capacitance. This Laplacian ensures that the currents between the two nodes ($A$ and $B$) flow in opposite directions, i.e., $I_{AB} = I_{BA}$, resulting in non-reciprocal impedance with current conversion. A unit cell of the circuit in Ref.~\cite{helbig_generalized_2020} is shown in Fig.~\ref{Fig_NH}a3, and includes an INIC (red box). 
Assembling the unit cells into a circuit array realizes the $k$-space circuit Laplacian:
\begin{equation}
	J(k)/i\omega = \epsilon_0 (\omega) \sigma_0 + (\nu(\omega) + r \cos(k))\sigma_x + (r \sin(k) - i \gamma) \sigma_y.
\end{equation}
When $\gamma = 0$, the Laplacian describes a Hermitian SSH circuit, with $\nu (\omega)$ and $r$ representing the reciprocal intra- and inter-cell coupling admittances, respectively. When $\gamma \neq 0$, the effective intra-cell coupling becomes non-reciprocal, leading to a non-Hermitian SSH circuit. All voltages localize at the left (right) boundary under OBCs when the driving frequency $f = \omega/2 \pi$ is below (above) the critical frequency [Left and Right plots in Fig.~\ref{Fig_NH}a4]. At the critical frequency, no localized skin modes are present.

Although a single asymmetric coupling within a unit cell may intuitively suggest boundary localization in the direction of asymmetry, this may not be true due to destructive interference~\cite{xue_topologically_2024}. Indeed, the cumulative effect of multiple asymmetric couplings with opposing decay profiles remains intriguing. Rafi-Ul-Islam et al.~\cite{rafi-ul-islam_unconventional_2022} investigated circuit lattices with fully non-reciprocal couplings [Fig.~\ref{Fig_NH}b1], with unit cells comprising four non-reciprocal nodes [Fig.~\ref{Fig_NH}b2]. Interestingly, the NHSE in their dimer SSH circuit with both intra- and inter-cell non-reciprocal couplings can vanish when the overall inverse localization length reaches a critical decay threshold, given by
\begin{equation}
	\kappa = -\frac{1}{2} \ln\abs{\frac{(C_1 - C_{n1})(C_2 + C_{n2})}{(C_1 + C_{n1})(C_2 - C_{n2})}},
\end{equation}
where $\kappa$ depends on both non-reciprocal capacitive parameters $C_{n1}$ and $C_{n2}$, and becomes zero when $C_{n1} = \frac{C_1 C_{n2}}{C_2}$. The localization boundary criterion remains governed by the sign of the overall inverse decay $\kappa$, consistent with established conventions. For example, in Fig.~\ref{Fig_NH}b3, the darker colors represent node voltage peaks that shift from one boundary to another (node 1 to 40) when $C_{n1}$ goes from below to above the critical value.

In a related vein, Shen et al.~\cite{shen_energy_2024} conducted a circuit experiment involving multiple asymmetric hoppings in coupled 1D SSH chains [Fig.~\ref{Fig_NH}c1]. They extended this setup by connecting multiple 1D SSH chains to form a spherical geometry, paving the framework for studying the interplay between band topology and real-space lattice topology~\cite{shang_observation_2024}. While localization behavior in various lattice geometries, including kagome and honeycomb lattices~\cite{zhang_finite-admittance_2021,huang_zero-admittance_2023}, has been widely studied, in the spherical model [Fig.~\ref{Fig_NH}c2, c3], localization occurs only when a defect is introduced [pointed by red arrow in Fig.~\ref{Fig_NH}c4]. Additionally, Jiang et al.~\cite{jiang_multiple_2024} demonstrated how circuits with multiple non-reciprocal couplings can give rise to corner modes in 2D circuits.

\subsubsection{Circuit demonstrations of higher-dimensional NHSE phenomena}

Typically, the NHSE drags all bulk states toward a boundary~\cite{okuma_non-hermitian_2023}. However, in two or more dimensions, Zou et al.~\cite{zou_observation_2021} demonstrated that the NHSE can selectively pump topological modes while leaving the bulk unaffected, in what is known as the hybrid higher-order skin-topological (ST) effect~\cite{lee_hybrid_2019}. The idea is that, in a multi-dimensional lattice, it is possible for topological modes under mixed boundary conditions to possess a spectral point gap, but not the non-topological modes. Further opening up the remaining boundary(ies) will then result in the NHSE only for the topological modes. In Ref.~\cite{zou_observation_2021}, the 2D network consists of a unit cell with four nodes, with asymmetric intra-cell couplings implemented using INICs [Fig.~\ref{Fig_NH}d1]. With suitable parameter tuning, the 2D and 3D circuit implementations [Figs.~\ref{Fig_NH}d2 and d3] exhibit skin-topological (ST) and skin-skin-skin (SSS) effects, respectively. These hybrid skin-topological modes arise from the interplay between topological and skin modes, and cannot exist without either. Notably, these ST modes can exhibit localized skin behavior despite the absence of net non-reciprocity, since the reciprocity is spontaneously broken by the topological modes taking unequal amplitudes in the two sub-lattices.

The hybrid skin-topological effect~\cite{lee_hybrid_2019} has also inspired a scheme for switching on or off the NHSE by toggling the topological character of the perpendicular sub-chains, as was first proposed in the cold atom context~\cite{li_topological_2020}. When the switch is activated by adjusting the loss parameter, non-reciprocal pumping prevails, driven by non-reciprocity introduced by the switch. Zhang et al.~\cite{zhang_electrical_2023} experimentally demonstrated this topological switching using their 3D circuit [Fig.~\ref{Fig_NH}e1]. The voltage dynamics across four different unit cells revealed distinct eigenmode distributions, such as skin-topo-skin and skin-topo-topo modes [Fig.~\ref{Fig_NH}e2, e3]. The absence or presence of skin modes corresponds to trivial or non-trivial topological modes, which are tuned through relays that switch resistors on or off between specific nodes.

The interplay between topology and the NHSE was further explored by Tang et al.~\cite{tang_competition_2023} through an experimental implementation of a rhombus honeycomb lattice electrical circuit [Fig.~\ref{Fig_NH}f1]. By tuning the non-reciprocity (controlled by $C_2$ in their design), the topologically localized corner modes at the acute angles of the rhombus lattice [Fig.~\ref{Fig_NH}f2] are significantly dragged through the bulk, allowing the NHSE to control the topological modes [Fig.~\ref{Fig_NH}f3, f4].

\subsubsection{Circuit demonstrations of interacting NHSE phenomena}

Despite being classical systems, electrical circuits can also simulate few-body interaction dynamics through mappings between many-body one-dimensional models and single-body multi-dimensional models~\cite{lee_many-body_2021,olekhno_topological_2020,gorlach2017topological,poddubny2023interaction}. Besides their practical value, such mappings can also lend new physical insights~\cite{shen_non-hermitian_2022,suthar_non-hermitian_2022,yang2024non}, such as the interpretation of two-body repulsion as 2-dimensional boundary effects, and 3-body non-Hermitian interactions as feedback mechanisms on a 2-dimensional non-orthogonal non-Hermitian lattice whose GBZ description is notoriously subtle~\cite{jiang_dimensional_2023}. In particular, effective boundaries can be defined in the Hilbert space even when the lattice has PBCs along each direction. Reverse implementations of such mappings have also enabled small interacting quantum systems to simulate the physics in far larger multi-dimensional lattices~\cite{koh2022simulation,koh_realization_2024,koh_interacting_2025}.

Zhang et al.~\cite{zhang_observation_2022} designed a circuit based on mapping the eigenstates of their strongly correlated non-Hermitian few-body system [Fig.~\ref{Fig_NH}g1]. Through impedance measurements, they observed the simulated aggregation of bosonic clusters in their circuit. The impedance resonances in the spatial circuit [Fig.~\ref{Fig_NH}g2] correspond to eigenstate aggregation in the corresponding Hilbert space [Fig.~\ref{Fig_NH}g3]. This realization serves as a compelling example of the feasibility of topolectrical circuits to model many-body interactions in a very accessible manner.

\subsubsection{Implementing NHSE without non-reciprocal couplings}
\label{sec:NHSE:reciprocal}

Even though paradigmatic NHSE models, i.e., the Hatano-Nelson model~\cite{hatano_localization_1996} almost always contain non-reciprocal (asymmetric) couplings, realizing the NHSE itself does not necessarily require asymmetric couplings. In the so-called reciprocal non-Hermitian skin effect, a reciprocal system is designed such that it contains subsystems harboring equal and opposite NHSE, such that NHSE accumulation occurs in specific subsectors despite the system being reciprocal on the whole. This was first demonstrated by Hofmann et al.~\cite{hofmann_reciprocal_2020}, who measured eigenmode localization at the edges of their 2D circuit [Fig.~\ref{Fig_NH}h1]. A diagonal imaginary hopping within the square plaquette makes the Hamiltonian globally non-Hermitian but still reciprocal; with a basis transform, it can be shown that it contains ``hidden'' equal and opposite NHSE chains at different transverse momentum sectors or effective flux. This imaginary hopping---realized as a resistor in their model [Fig.~\ref{Fig_NH}h2]---results in different hopping probabilities in the presence of this effective flux, resulting in unequal amplitudes in opposite directions [Fig.~\ref{Fig_NH}h3].

In two dimensions, a reciprocal 2D lattice can also exhibit nontrivial system size-dependent NHSE in the presence of higher-rank chirality~\cite{li_large-chiral-number_2023}, characterized by a non-conserved charge current~\cite{dubinkin_higher_2024}. First demonstrated in Ref.~\cite{li_large-chiral-number_2023}, a higher-rank chiral mode (rank-2) was also demonstrated by Zhu et al.~\cite{zhu_higher_2023} through gain/loss couplings that create an unbalanced chirality in the unidirectional chiral mode flow [Fig.~\ref{Fig_NH}i1]. This imbalance causes one specific mode to amplify and hence outlast others, leading to eventual mode localization. The long-lived rank-2 chiral mode dominates the system's long-time dynamics and is observable in momentum-resolved measurements [Figs.~\ref{Fig_NH}i2, i3]. Remarkably, momentum-dependent input excitations lead to mode accumulation at specific driven locations in the lattice [Fig.~\ref{Fig_NH}i4]. This enables a versatile implementation of NHSE along a desired direction and position in a fully passive and reciprocal lattice, distinguishing it from the reciprocal skin effect presented in Ref.~\cite{hofmann_reciprocal_2020}. (Even in the absence of gain/loss and non-reciprocity in the lattice bulk, non-Hermiticity can emerge through the interaction Hamiltonian in an electromagnetically engineered metamaterial that is inversely designed from a TE lattice~\cite{maopeng_gainloss-free_2024}. See also Ref.~\cite{li_observation_2025_2}, where the NHSE is realized in reciprocal lattices via gauge fields without gain or loss.)

\subsubsection{Demonstrating unconventional NHSE phenomena with long-range couplings}

The extreme versatility of electrical circuit connections, which can directly couple any two desired nodes independent of their physical distance, makes them ideal for implementing intrinsically long-ranged models~\cite{yin_observation_2025}. Since the NHSE is already highly non-local in that it leads to dramatic state accumulation in a spatially-distant edge, it is particularly susceptible to the effects of long-ranged hoppings that can change the overall real-space topology of the lattice. For instance, a single non-reciprocal impurity hopping in an otherwise homogeneous Hatano-Nelson chain is known to lead to enigmatic scale-free localization~\cite{li_impurity_2021,su_observation_2023,guo_accumulation_2023}. 

Recently, Guo et al.~\cite{guo_scale-tailored_2024} introduced the new notion of scale-tailored localization (STL), distinct from conventional NHSE or Anderson localization~\cite{yilmaz_scale-free_2024}. The STL is a spectacular demonstration of how singularities in the skin localization length subtly interplays with hopping non-locality, as was implemented via unidirectional couplings with INIC circuit elements, and a long-range coupling controlled by switches [Fig.~\ref{Fig_NH}j1]. For this circuit, some of the admittance eigenvalues encircle the unit circle, while others are isolated within it [Fig.~\ref{Fig_NH}j2]. The spatial distribution of the corresponding STL eigenstates exhibits varying localization lengths that depend on the long-range coupling distance [Fig.~\ref{Fig_NH}j3].

\subsection{Circuit applications for non-Hermitian sensing}

Since the NHSE is a dramatic phenomenon that arises from merely changing the boundary couplings, it is symptomatic of extreme sensitivity to spatial perturbations. Below, we highlight some experiments that attempt to point towards potential sensing applications in electrical circuit settings.

\begin{figure}[ht!]
	\centering
	\includegraphics[width=\linewidth]{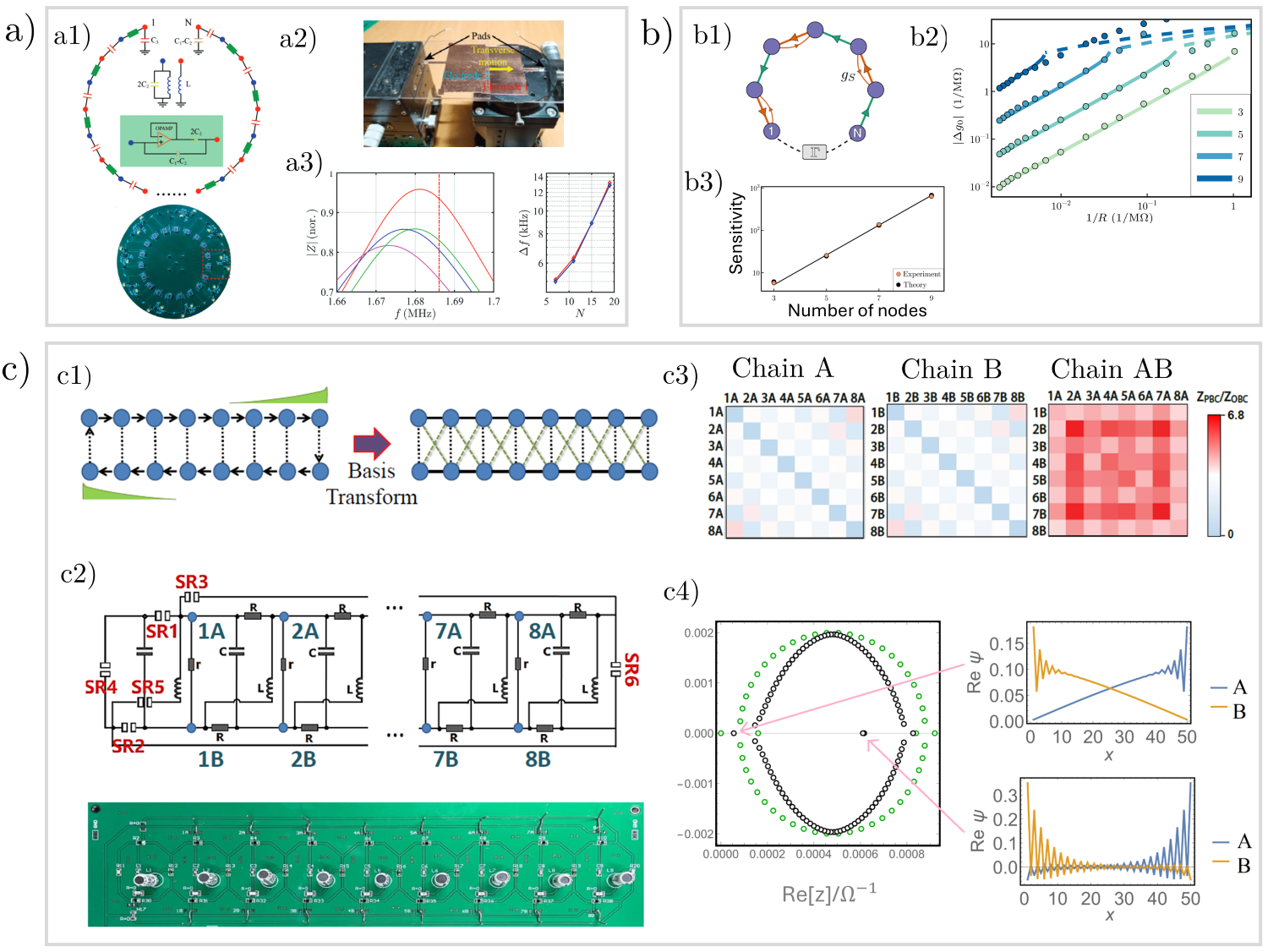}
	\caption{\textbf{Non-Hermitian sensing applications with electrical circuits.}  
		\textbf{(a)} Non-Hermitian topological sensor realized in a circuit with adjustable boundary perturbations and circuit size, with size-dependent shifts in the impedance peak frequency as shown in a3~\cite{yuan_nonhermitian_2023}. [Adapted with permission from Yuan et al., Advanced Science, 10(19), 2301128 (2023). Copyright 2023 licensed under CC BY 4.0.]
		\textbf{(b)} Non-Hermitian ohmmeter exhibiting exponential sensitivity to large resistance measurements. The sensitivity increases exponentially as the number of nodes increases, as shown in b3~\cite{konye_non-hermitian_2024}. [Adapted with permission from K\"onye et al., Phys. Rev. Applied 22, L031001 (2024). Copyright 2024 by the American Physical Society.]
		\textbf{(c)} Non-local impedance response in a fully passive circuit~\cite{zhang_observation_2024}. 
		\textbf{(c1)} The NHSE in two weakly coupled Hatano-Nelson 1D lattices may not cancel off despite zero net reciprocity, and can be basis-transformed into a purely lossy reciprocal model.  
		\textbf{(c2)} The corresponding purely resistive electrical circuit with switches (SR1 to SR6) that controls the boundary conditions.  
		\textbf{(c3)} Ratio of the PBC vs. OBC impedances between nodes in the two physical chains, with substantially different PBC and OBC impedances for inter-chain impedances (red, right plot).
		\textbf{(c4)} Robust isolated boundary modes that emerge only in the presence of parasitic resistances and sufficiently large system length. [(c1,c2,c3,c4) adapted with permission from Zhang et al., SciPost Phys. 16, 002 (2024). Copyright 2024 licensed under CC BY 4.0.]}
	\label{Fig_NLimpedance}
\end{figure}

The primary interest in the NHSE stems from its strong dependence on local perturbations, such as a coupling between the two ends of a chain, which can drastically alter the OBC vs. PBC properties. Furthermore, a non-Hermitian topological lattice exhibits extreme sensitivity to its size due to the exponential growth of NHSE states, as formulated theoretically by Budich and Bergholtz~\cite{budich_non-hermitian_2020}, and popularized as non-Hermitian topological sensors~\cite{rafi-ul-islam_saturation_2024}. This phenomenon was experimentally demonstrated by Yuan et al.~\cite{yuan_nonhermitian_2023} [Fig.~\ref{Fig_NLimpedance}a1]. The shift in resonance peak frequency occurs through two mechanisms: (i) control of the boundary coupling capacitance through the displacement and rotation of the copper electrodes [Fig.~\ref{Fig_NLimpedance}a2], and (ii) variations in the number of nodes in the circuit. The impedance resonances exhibit exponential behavior in response to circuit size $N$ [Fig.~\ref{Fig_NLimpedance}a3].

Based on the exponential response to boundary perturbations in non-Hermitian circuits, K\"onye et al.~\cite{konye_non-hermitian_2024} designed a non-Hermitian topological ohmmeter with reciprocal coupling between the first and last terminals (reminiscent to Ref.~\cite{yuan_nonhermitian_2023}), utilizing it as a sensor to precisely measure large resistances ($\Gamma$ in Fig.~\ref{Fig_NLimpedance}b1), typically in the mega-ohm range. The measurement of large resistances often suffers from reduced precision due to significant losses. However, their multi-terminal non-Hermitian ohmmeter circuit demonstrated exponentially increasing sensitivity [Fig.~\ref{Fig_NLimpedance}b2] with the number of terminals, comparing favorably with standard measurement techniques [Fig.~\ref{Fig_NLimpedance}b3].

Enhanced response can also be achieved through various mechanisms, such as multiple non-local long-range asymmetric couplings, which lead to an $N$th-order sensitive response, where $N$ represents the number of nodes, as introduced in Refs.~\cite{zhang_true_2024,chen_ultra-sensitivity_2024}. Ref.~\cite{deng_ultrasensitive_2024} also presents an integrated circuit sensor demonstrating an exponential response with respect to the device size, based on higher-order non-Hermitian topology.

In the previous non-Hermitian sensing experiments, the underlying mechanism is directional amplification, which results in a strongly non-local response. Indeed, exponential sensitivity intuitively requires a non-reciprocal non-Hermitian bulk, achieved through asymmetric couplings~\cite{kong_manipulating_2024}, which necessitate active elements such as op-amps in INICs~\cite{jiang_tunable_2024,liao_non-hermitian_2024}. However, Zhang et al.~\cite{zhang_observation_2024} presented a fully passive circuit demonstrating a highly distinct impedance response dependent on the presence of a boundary coupling, even though the circuit uses only resistors as the non-Hermitian components.

The non-local impedance response in this passive circuit is due to the presence of ``hidden'' coupled Hatano-Nelson chains with opposite amplifications, as seen through a basis rotation [Fig.~\ref{Fig_NLimpedance}c1]. After the basis transform, the effective momentum space circuit Laplacian $\tilde{J}(k)$ at the resonant frequency is expressed as
\begin{equation}
\tilde{J}(k)|_{\omega=\omega_0} =\left(\begin{array}{cc}
		2 \omega_0 C e^{i k}+\xi(k) & -r^{-1} \\
		-r^{-1} & 2 \omega_0 C e^{-i k}+\xi(k)
	\end{array}\right),
\end{equation}
where $\xi(k) = r^{-1} +2R^{-1}-2(R^{-1}+\omega_0 C) \cos(k)$. Here, $\xi(k)$ represents the reciprocal part of each chain, while $2 \omega_0 C e^{\pm i k}$ denotes the asymmetric contributions. The two chains, coupled by $-r^{-1}$, effectively experience opposite NHSE due to unbalanced gain/loss couplings. The boundary conditions of the corresponding circuit [Fig.~\ref{Fig_NLimpedance}c2] are controlled via switches, appropriately configured for desired behavior.

Experimentally, the circuit's response to non-local perturbations is characterized by the ratio of two-point impedance values under PBCs vs. OBCs: $Z_\text{PBC}/Z_\text{OBC}$. While this impedance ratio across all two-point configurations within individual chains remains approximately equal to 1 [the two light blue plots on the left in Fig.~\ref{Fig_NLimpedance}c3], the impedance across inter-ladder pairs deviates significantly, with the ratio hovering around 6 [red in the right plot in Fig.~\ref{Fig_NLimpedance}c3]. This indicates substantial non-local impedance response to the boundary coupling, which is interestingly obtained in a purely resistive circuit, without using any amplificative components such as op-amps.

While this setup may seem to be just another demonstration of the reciprocal NHSE~\cite{hofmann_reciprocal_2020} (see Fig.~\ref{Fig_NH}h in Section~\ref{sec:NHSE:reciprocal}), the effective inter-chain $-1/r$ coupling is set to be relatively weak compared to the intra-chain terms, such that the NHSE from both chains does not simply cancel in the intended experimental setup. However, due to the exponential sensitivity of the NHSE, they still do cancel for sufficiently long circuits, as a variation of the critical NHSE~\cite{li_critical_2020,qin_universal_2023,rafi-ul-islam_critical_2022,siu_terminal-coupling_2023} (refer to the discussion related to Fig.~\ref{Fig_NH}b in Section~\ref{sec:NHSE:multiple}). In this model, the circuit exhibits curious size-dependent isolated modes \emph{only} in the presence of parasitic resistances [Fig.~\ref{Fig_NLimpedance}c4], when the OBC circuit is sufficiently long. Their spatial profiles are boundary-localized in a unique manner, as shown in the two rightmost plots in Fig.~\ref{Fig_NLimpedance}c4. Despite being robust against perturbations, such isolated modes do not appear to be protected by any conventional topological invariant, appearing at certain ranges of system sizes.

\section{Circuit implementations of condensed-matter phenomena}

Below, we highlight other works on topolectrical circuits, mostly experimental, that are centered around more traditional (mostly Hermitian) condensed matter phenomena. In this review, heavier emphasis is placed on the simulation of phenomena that are not easily achieved in conventional quantum matter.

\subsection{Non-linear phenomena in topolectrical circuits}

Introducing non-linear dynamics into a system reveals intriguing phenomena that are unattainable in linear systems, such as higher harmonic generation~\cite{lapine_colloquium_2014,marquie_nonlinear_1995,kengne_ginzburglandau_2022}, soliton wave propagation~\cite{li_higher-order_2019,hohmann_observation_2023,jana_harnessing_2025,li_observation_2024}, non-linear bands~\cite{tuloup2020nonlinearity}, non-linear interface modes~\cite{tang_strongly_2023} as well as various non-Hermitian interplays~\cite{xia_nonlinear_2021,lang_non-hermitian_2021,many_manda_insensitive_2024,veenstra_non-reciprocal_2024,akgun2024deterministic,dai2024non,sone2025transition}. Specifically, in the presence of non-linear interactions, coherent collective dynamics can emerge among the autonomous units hosting non-linear oscillators. Of particular interest is the incorporation of a topological mechanism, which facilitates self-organized, protected boundary oscillations~\cite{sone_topological_2022,xia_nonlinear_2021,tang_strongly_2023}. A common approach to realizing non-linear topological phenomena involves arranging non-linear oscillatory elements in a topological configuration, such as by dimerizing an array of oscillators in deference to the well-known SSH configuration.

\begin{figure*}[ht!]
	\centering
	\includegraphics[width=\textwidth]{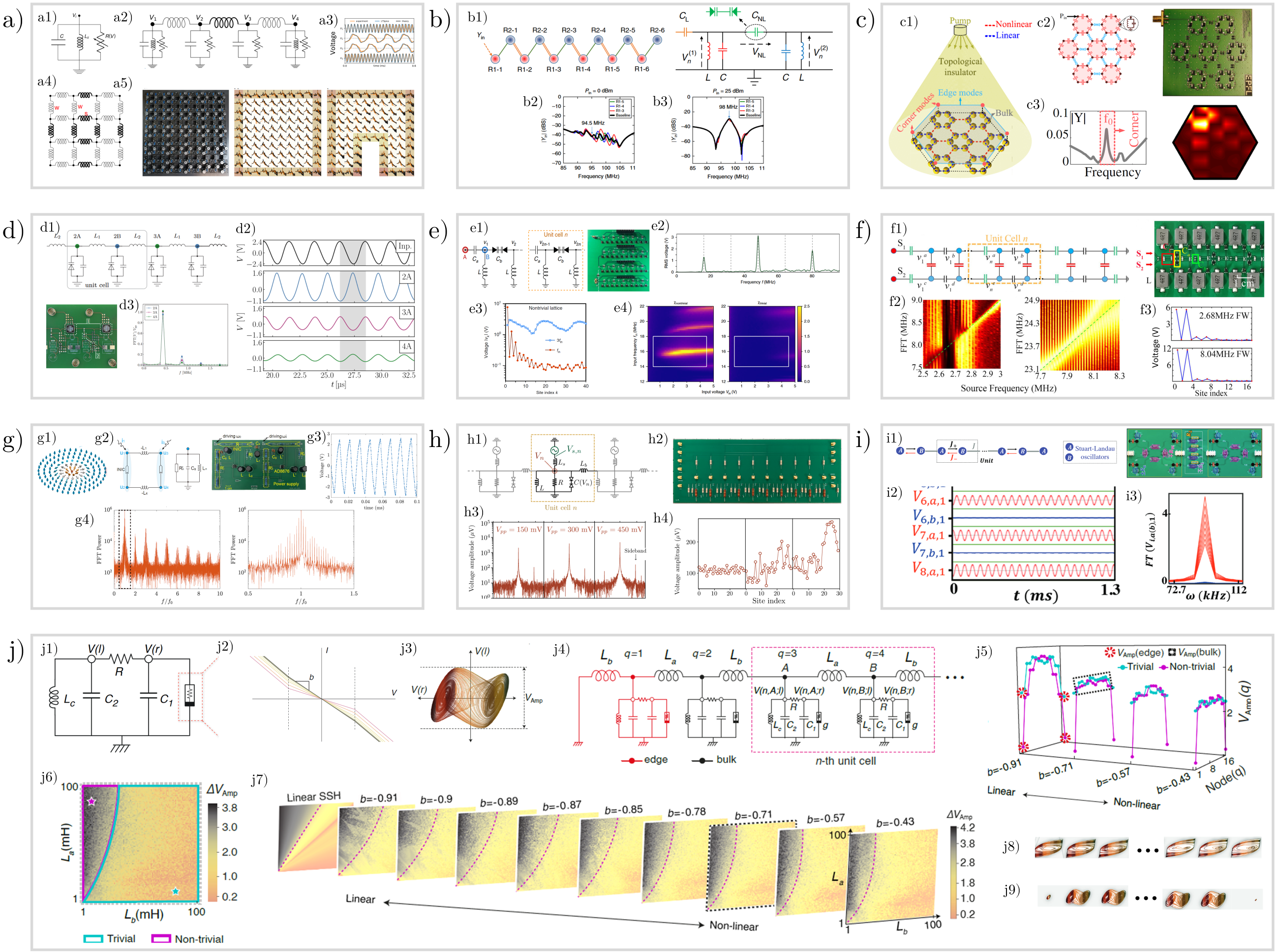}
	\caption{\textbf{Realization of non-linear phenomena in topolectrical circuits.}  
		\textbf{(a)} Active topolectrical circuits with autonomous non-linear on-site oscillators exhibit synchronized oscillations at sites corresponding to topological modes~\cite{kotwal_active_2021}. [Adapted with permission from Kotwal et al., PNAS, 118(32), e2106411118 (2021). Copyright 2021 licensed under CC BY 4.0.]
		\textbf{(b)} Non-linear transmission line containing voltage-dependent intra-cell capacitors. A topological phase transition (the peak at 98 MHz in b3) occurs with sufficiently strong transmitted signal intensity~\cite{hadad_self-induced_2018}. [Adapted with permission from Hadad et al., Nat Electron 1, 178–182 (2018). Copyright 2018 by Springer Nature.]
		\textbf{(c)} Second-order non-linear circuit with topological corner modes activated by the intensity of the input signal~\cite{zangeneh-nejad_nonlinear_2019}. [Adapted with permission from Zangeneh-Nejad and Fleury, Phys. Rev. Lett. 123, 053902 (2019). Copyright 2019 by the American Physical Society.]
		\textbf{(d)} Generation of cnoidal solitonic waves in a topological circuit with non-linear grounded capacitors~\cite{hohmann_observation_2023}. [Adapted with permission from Hohmann et al., Phys. Rev. Research 5, L012041 (2023). Copyright 2023 licensed under CC BY 4.0.]
		\textbf{(e)} Topological enhancement of third-harmonic generation in a non-linear transmission line with voltage-dependent intra-cell capacitors~\cite{wang_topologically_2019}. [Adapted with permission from Wang et al., Nat Commun 10, 1102 (2019). Copyright 2019 licensed under CC BY 4.0.]
		\textbf{(f)} Topologically enhanced third-harmonic generation in a ladder circuit with double-edge localization~\cite{hu_observation_2024}. [Adapted with permission from Hu et al., Commun Phys 7, 204 (2024). Copyright 2024 licensed under CC BY 4.0.]
		\textbf{(g)} Observation of frequency combs induced by the non-linear behavior of strongly pumped operational amplifiers~\cite{yang_simulating_2024}. [Adapted with permission from Yang et al., arXiv:2410.19598 (2024). Copyright 2024 licensed under CC BY 4.0.]
		\textbf{(h)} Non-linearity-induced non-Hermitian skin effect, realized through Kerr-type non-linearity implemented by back-to-back onsite varactors~\cite{lo_switchable_2024}. [Adapted with permission from Lo et al., arXiv:2411.13841 (2024). Copyright 2024 licensed under CC BY 4.0.]
		\textbf{(i)} Global synchronization due to the non-Hermitian skin effect in a non-linear circuit with on-site Stuart-Landau oscillators~\cite{zhang_nonhermitian_2024}. [Adapted with permission from Zhang et al., Advanced Science, 12(2), 2408460 (2025). Copyright 2025 licensed under CC BY 4.0.] 
		\textbf{(j)} Robust edge suppression and topologically protected chaotic oscillations in a lattice of Chua circuits. The non-linearity can be controlled by adjusting the current-voltage slope in the Chua diode~\cite{sahin_protected_2024}. [Adapted with permission from Sahin et al., arXiv:2411.07522 (2024). Copyright 2024 licensed under CC BY 4.0.]}
	\label{Fig_NonLinear}
\end{figure*}

In electrical circuits, non-linear oscillators necessarily involve active components such as op-amps, and typically do not require external driving. Kotwal et al.~\cite{kotwal_active_2021} demonstrated an archetypal topological circuit implementation hosting non-linear van der Pol (vdP) oscillators. The dynamics of a single oscillator with coordinate $x_i$ is governed by
\begin{equation}
	\ddot{x}_i - \epsilon (1-x_i^2)\dot{x}_i + x_i = 0,
\end{equation}
where $\epsilon=\alpha \sqrt{L/C}$, $t=\sqrt{LC}\hat{t}$, $\dot{x}_i = dx_i/d\hat{t}$, and $t$ is physical time. When the tunable real positive parameter $\alpha$ is zero, the vdP differential equation reduces to that of a simple harmonic oscillator, representing a linear system. However, vdP oscillators exhibit non-linear oscillations for $\alpha \neq 0$, driven by a negative non-linear resistance that can be tuned via $\alpha$. In practice, this negative resistance in a unit [Fig.~\ref{Fig_NonLinear}a1] is realized using a Chua diode, which exhibits non-linear current-voltage characteristics. A 1D array of these oscillators [Fig.~\ref{Fig_NonLinear}a2], arranged in a topological dimer configuration, exhibits distinct oscillations at the edge nodes, reminiscent of topological edge modes [Fig.~\ref{Fig_NonLinear}a3]. Extending this concept to a 2D network [Fig.~\ref{Fig_NonLinear}a4], the vdP oscillators display chiral-like edge oscillations that remain robust even in the presence of edge defects [Fig.~\ref{Fig_NonLinear}a5]. These circuits, composed of onsite autonomous units, are referred to as active topolectrical circuits.

In addition to having autonomous units serving as on-site potentials in a topological circuit array, it is also possible to use non-linear components, such as diodes or voltage-dependent capacitors (varactors), for inter- or intra-cell coupling. Hadad et al.~\cite{hadad_self-induced_2018} realized a non-linear SSH circuit by alternating linear and non-linear capacitances [Fig.~\ref{Fig_NonLinear}b1]. The non-linear capacitance is implemented using two varactor diodes connected in a back-to-back configuration, with the capacitance dependent on the voltage across the diodes' terminals:
\begin{equation}
	\nu (V_n) = \frac{\nu_0 - \nu_\infty}{1+V_n/V_0}+\nu_\infty,
\end{equation}
where $\nu(V_n)$ represents the intra-cell coupling admittance that can be non-linearly controlled by the capacitor voltage $V_n$, and $\nu_0$, $\nu_\infty$, and $V_0$ are the characteristic constants of the coupling~\cite{hadad_self-induced_2018}. This setup makes the non-linear coupling ($\nu(V_n)$) sensitive to the intensity of the driving signal. As the signal intensity increases, a topological transition is induced, leading to the emergence of topological edge modes. This transition was observed through admittance versus frequency measurements. Below the threshold signal intensity, no distinct resonance is detected [Fig.~\ref{Fig_NonLinear}b2]. However, when the signal intensity exceeds the threshold, a pronounced admittance peak appears at the expected resonant frequency [Fig.~\ref{Fig_NonLinear}b3].

A key aspect of such non-linear topolectrical circuit realizations is the ability to achieve a topological phase transition through the excitation intensity. This dependence eliminates the need for having separate lattice configurations for trivial or non-trivial phases, and allows access to robust topological modes through external pumping [Fig.~\ref{Fig_NonLinear}c1]. Zangeneh-Nejad and Fleury~\cite{zangeneh-nejad_nonlinear_2019} subsequently realized a second-order non-linear TE [Fig.~\ref{Fig_NonLinear}c2] that emulates second-order topological corner modes when the input intensity ($P_\text{in}$) exceeds a certain threshold. This implementation employs non-linear varactor diodes, similar to Ref.~\cite{hadad_self-induced_2018}. For instance, when $P_\text{in}$ is 25dBm, an admittance resonance occurs, corresponding to the second-order topological 0-dimensional corner mode [Fig.~\ref{Fig_NonLinear}c3].

Moving away from using varactors as non-linear coupling components, Hohmann et al.~\cite{hohmann_observation_2023} employed varactors to create non-linear on-site potentials [Fig.~\ref{Fig_NonLinear}d1]. This approach led to qualitatively different outcomes: injected sinusoidal waves were distorted into localized cnoidal waves---periodic, soliton-like waves [Fig.~\ref{Fig_NonLinear}d2]. The localization behavior is attributed to topological exponential boundary localization, evident in the root-mean-square amplitudes decaying toward the bulk, demonstrative of the interplay between topological and solitonic localization. Additionally, the non-linearity generates higher harmonics relative to the circuit's driving frequency, although the amplitudes of these harmonics decrease with increasing harmonic order [Fig.~\ref{Fig_NonLinear}d3].

Using back-to-back varactors as well, Wang et al.~\cite{wang_topologically_2019} introduced a non-linear transmission line circuit in which varactors were employed as intra-cell non-linear coupling components [Fig.~\ref{Fig_NonLinear}e1]. This circuit exhibits third-harmonic signals [Fig.~\ref{Fig_NonLinear}e2] that are much larger than the first harmonic. Interestingly, the higher-harmonic signal [blue in Fig.~\ref{Fig_NonLinear}e3] propagates through the bulk without diminishing in amplitude. Instead, its amplitude is significantly enhanced, surpassing the edge-localized first-harmonic signal [red in Fig.~\ref{Fig_NonLinear}e3]. Unlike in the setup of Ref.~\cite{hadad_self-induced_2018}, this enhanced higher-harmonic behavior is attributed to the traveling-wave nature of higher-harmonic modes, which can excite the entire circuit. In contrast, the first harmonic remains localized at the edge in the deep non-linear topologically non-trivial regime [Fig.~\ref{Fig_NonLinear}e4].

Hu et al.~\cite{hu_observation_2024} demonstrated greatly enhanced third-harmonic resonances in a mirror-stacked design consisting of two non-linear chains, each unit cell coupled by a linear capacitor [Fig.~\ref{Fig_NonLinear}f1]. This mirror-stacked design enables the realization of both fundamental harmonics and third harmonics due to the presence of multiple edge nodes with two distinct frequency localized states. Under single-source excitation, the circuit exhibits prominent third harmonic resonances [Fig.~\ref{Fig_NonLinear}f2], with identical voltage profiles in the first and third harmonics [Fig.~\ref{Fig_NonLinear}f3].

While most non-linearity-related phenomena arise from non-linear elements such as varactors, Yang et al.~\cite{yang_simulating_2024} observed a frequency comb generated by the non-linear behavior of op-amps under strong pumping. Frequency combs represent the proliferation of strong higher harmonics~\cite{picque_frequency_2019}. Yang et al.'s circuit was designed to simulate skyrmion [Fig.~\ref{Fig_NonLinear}g1] dynamics in the circuit unit within each node~\cite{tokura_magnetic_2021} [Fig.~\ref{Fig_NonLinear}g2]. Remarkably, distorted voltage oscillations resulting from op-amp saturation under a strong input signal gave rise to discrete, evenly spaced frequency combs [Fig.~\ref{Fig_NonLinear}g3]. This observation highlights how non-linear behavior can emerge from devices inherently designed for linear operation, offering a controlled method for introducing non-linearity in circuit designs.

Lo et al.~\cite{lo_switchable_2024} recently presented a non-linear circuit experiment demonstrating NHSE that emerges depending on the intensity of the driving amplitude. Interestingly, this behavior arises not from asymmetric couplings but from Kerr-type non-linear dynamics~\cite{yuce_nonlinear_2025,many_manda_skin_2024,longhi_modulational_2025}. The circuit comprises back-to-back onsite non-linear varactors [Fig.~\ref{Fig_NonLinear}h1] assembled into a circuit array inspired by the physics of Bogoliubov modes~\cite{bogoliubov1947theory} [Fig.~\ref{Fig_NonLinear}h2]. The non-linearity breaks pseudo-Hermiticity, leading to a spectral point gap that in turn gives rise to NHSE localization. The sideband modes [Fig.~\ref{Fig_NonLinear}h3], distinct from higher-order harmonics, exhibit localized profiles that depend on the auxiliary drive frequency. Due to the Kerr non-linearity, the NHSE appear only when the input signal amplitude exceeds a certain threshold [Fig.~\ref{Fig_NonLinear}h4]. This work highlights how non-linearity can induce the NHSE in classical electrical circuits.

The complex emergent dynamical behavior in non-linear systems can sometimes pose significant challenges for achieving global synchronization~\cite{acebron_kuramoto_2005}. However, Zhang et al.~\cite{zhang_nonhermitian_2024} recently demonstrated a Hatano-Nelson circuit hosting on-site non-linear Stuart-Landau oscillators [Fig.~\ref{Fig_NonLinear}i1]. Due to the asymmetric couplings in the Hatano-Nelson circuit, Hermiticity is broken, resulting in non-orthogonal eigenmodes. This facilitates more effective communication between modes, ultimately leading to global synchronization. This behavior manifests as collective voltage oscillations observed in the time domain [Fig.~\ref{Fig_NonLinear}i2] and in the steady-state voltage profile [Fig.~\ref{Fig_NonLinear}i3]. This study highlights how the NHSE can enable global synchronization in non-linear experimental circuits (also refer to Ref.~\cite{di_observation_2025}).

A cornerstone of many non-linear systems is their ability to support chaotic dynamics~\cite{lorenz1963deterministic,may_simple_1976,acebron_kuramoto_2005,sahin_protected_2024}. A paradigmatic chaotic system is the Chua oscillator which, unlike Stuart-Landau oscillators, exhibits irregular and chaotic orbits in some parameter regimes~\cite{kennedy_three_1993}. Mathematically described by the Lorenz system of differential equations, the Chua oscillator possesses a well-known circuit realization known as the Chua circuit~\cite{chua1992genesis,chua_tenyears_1994,chua_dynamic_1980}. It can be constructed with four passive components and one active component known as the Chua diode~\cite{ruy_fourelement_2008}[Fig.~\ref{Fig_NonLinear}j1], which provides the requisite non-linear negative resistance. This non-linearity is evident in its non-linear current-voltage ($IV$) characteristic [Fig.~\ref{Fig_NonLinear}j2], and typical chaotic behavior includes the famed double-scroll chaotic attractor [Fig.~\ref{Fig_NonLinear}j3].

Of particular interest is how such chaotic behavior interplays with lattice band structure dynamics, such as when multiple non-linear chaotic oscillators are coupled to form an array. In Ref.~\cite{sahin_protected_2024}, Sahin et al. investigated the consequences of coupling Chua oscillators in a topological SSH-like lattice structure [Fig.~\ref{Fig_NonLinear}j4]. Since band topology is only rigorously defined in linear systems, one might intuitively expect the non-linearity, and especially its chaotic dynamics, to erode the topological localization. However, it turns out that topological edge localization persists robustly well into the non-linear regime, with the amplitudes of the chaotic scroll being very different at the bulk and edge oscillators [Fig.~\ref{Fig_NonLinear}j5] \emph{only} in the black parameter space region [Fig.~\ref{Fig_NonLinear}j6]---the region extrapolated from the topological non-trivial parameter region defined in the linear limit. This extrapolation can be performed by tuning the $I$-$V$ characteristic parameter $b$ of the Chua diode [Fig.~\ref{Fig_NonLinear}j7], and explicitly showcases how the topological edge oscillation suppression robustly carries over from the linear into the non-linear regime. 

By externally perturbing the voltage oscillations by injecting current at the left edge node, one further observes that the chaotic oscillations, which are typically highly sensitive to external perturbations, persist robustly only when in the topologically non-trivial regime. In the regime connected to the trivial phase, the injected current reduces the chaotic double-scroll portraits to non-chaotic limit-cycle oscillations [Fig.~\ref{Fig_NonLinear}j8]. However, in the non-trivial phase, the same current injection only disturbs the edge node oscillations, leaving the bulk oscillations mostly intact [Fig.~\ref{Fig_NonLinear}j9]. This robustness is captured as a phenomenon termed ``topologically protected chaos'', where chaotic dynamics counterintuitively become more robust by the remnants of topological localization deep inside the non-linear regime~\cite{sahin_protected_2024}.

\subsection{Floquet circuits and the observation of temporal phenomena}

Typical realizations of tight-binding condensed matter phenomena map the lattice model onto a real space physical array, and are often met with challenges in implementing or maintaining a large number of lattice sites to within acceptance levels of error tolerance. One promising alternative approach is to instead realize the model in the frequency lattice, through a time-periodic driving approach known as Floquet engineering~\cite{rudner_anomalous_2013,traversa_generalized_2013,cayssol_floquet_2013,rodriguez2021low,sentef2015theory,lee2018floquet,kennes2018floquet,li2018realistic,oka_floquet_2019,wintersperger_realization_2020,rudner_band_2020,cai_non-hermitian_2024,harper_topology_2020,lee2020ultrafast,li2021quantized,qin2023light,qin2024light,yin_floquet_2022}. In analogy to the Fourier correspondence between a periodic Brillouin zone and its real-space lattice, a system that is periodically driven in time should also correspond to a lattice in frequency space. This powerful analogy enables the implementation of complicated hoppings and unit cells in terms of appropriately programmed temporal driving frequency components, all within a single (time-periodic) physical unit cell.

A key difference between Floquet-driven Hamiltonian systems and Floquet-driven electrical circuits is that in the latter (electrical circuits), the main physics is contained in the steady-state relationship between the current and the voltage, i.e., the circuit Laplacian $J$ that encapsulates Kirchhoff's law $I =JV$ (refer to Section~\ref{sec:laplacian}). As such, impedance measurements probe the properties of the Laplacian $J$, rather than the time-evolution generator $H$. As explained in Ref.~\cite{stegmaier_topological_2024}, this fundamental difference is reflected in the frequency dependence of the Floquet operator: while ordinary Floquet Hamiltonians can only contain a linear ``background'' potential in the frequency lattice~\cite{martin_topological_2017,sridhar_quantized_2024}, a Floquet Laplacian can depend on the frequency in arbitrarily complicated manner, depending on the frequency dependencies of its constituent components, i.e., $i\omega C$ or $1/i\omega L$ for capacitive and inductive components. As such, Floquet circuit Laplacians can be engineered to possess a variety of background ``potentials'' in the frequency lattice, such that one can also probe impurity or edge phenomena, i.e., topological edge states in appropriately designed temporally driven electrical circuits.

\begin{figure}%[ht!]
	\centering
	\includegraphics[width=\linewidth]{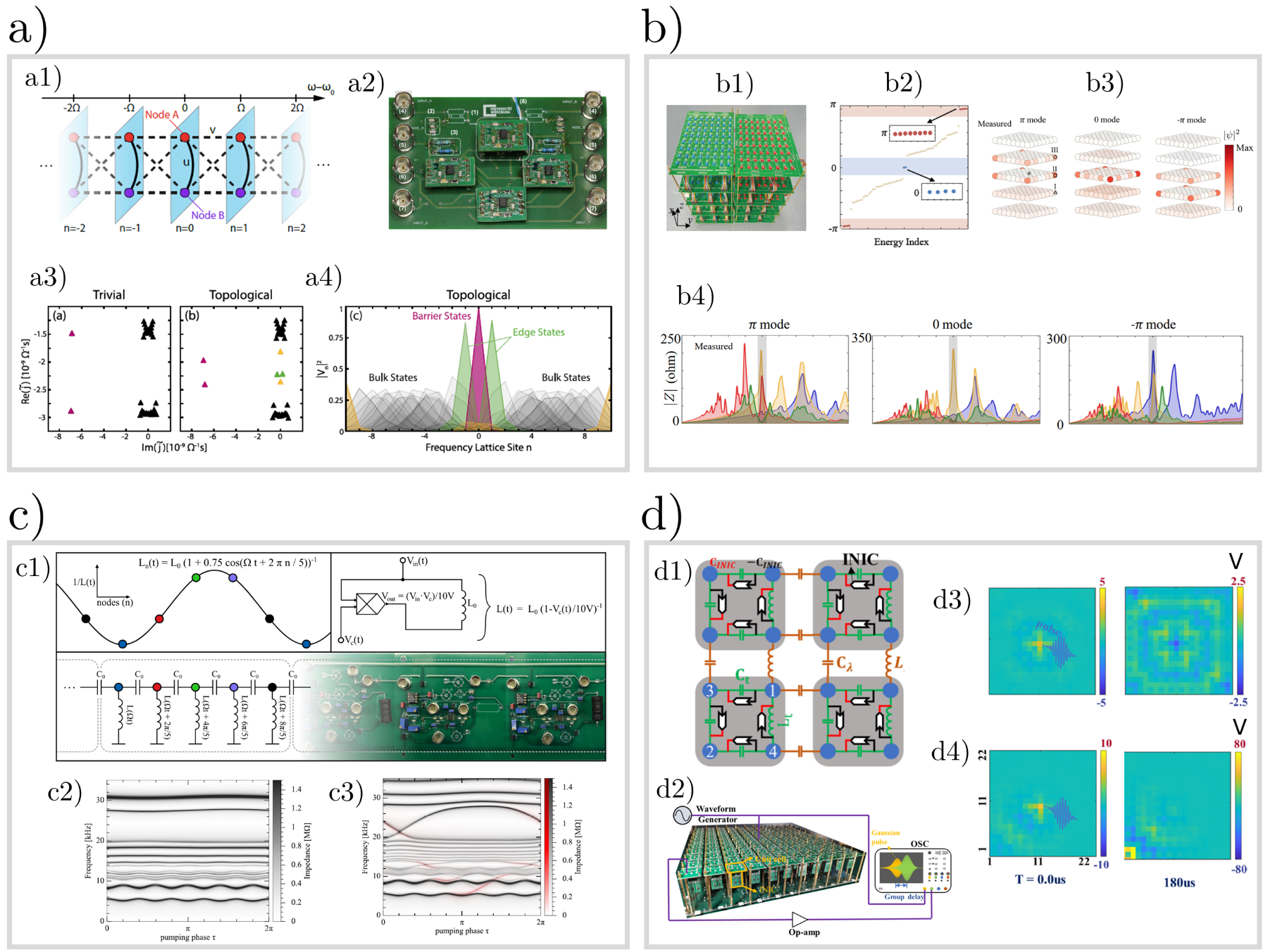}
	\caption{\textbf{Circuit experiments for Floquet phenomena and temporal dynamics.}
		\textbf{(a)} Floquet circuit that showcases topological boundary accumulation at background inhomogeneities of the Floquet Laplacian in frequency space, with time-modulation physically achieved through the use of an analog multiplier and an input signal~\cite{stegmaier_topological_2024}. [Adapted with permission from Stegmaier et al., arXiv:2407.10191 (2024). Copyright 2024 licensed under CC BY 4.0.]
		\textbf{(b)} Observation of time-dislocated Floquet topological modes in a 3D electrical circuit, with the third (inter-layer) dimension representing the frequency lattice~\cite{zhang_Floquet_observation_2024}. [Adapted with permission from Zhang et al., Nat Commun 16, 2050 (2025). Copyright 2025 licensed under CC BY 4.0.] 
		\textbf{(c)} Demonstration of topological AAH spectral pumping through dynamical voltage-controlled inductors~\cite{stegmaier_realizing_2024}. [Adapted with permission from Stegmaier et al., Phys. Rev. Research 6, 023010 (2024). Copyright 2024 licensed under CC BY 4.0.]
		\textbf{(d)} Observation of the temporal corner skin dynamics from the higher-order non-Hermitian skin effect in a 2D circuit~\cite{wu_evidencing_2023}. [Adapted with permission from Wu et al., Phys. Rev. B 107, 064307 (2023). Copyright 2023 by the American Physical Society.]}
	\label{Fig_Floquet}
\end{figure}

Stegmaier et al.~\cite{stegmaier_topological_2024} designed a circuit with two logical nodes that assumes the topological SSH form in the frequency domain [Fig.~\ref{Fig_Floquet}a1]. By employing a combination of AD633 analog multipliers and capacitors  [Fig.~\ref{Fig_Floquet}a2], effectively time-varying capacitors were achieved, modulated at far higher frequencies (order of MHz) than what is possible with mechanical means. To recreate an SSH model in frequency space, the inter- and intra-node capacitances are time-modulated such that their temporal functional dependence resembles that of the momentum dependence of the conventional SSH model. Notably, the Floquet driving protocol is such that the effective background potential diverges at the zeroth ``site'' in the frequency lattice, such that it acts as a strong ``barrier''. As conspicuously measured, topology boundary states exist at frequencies very near these barriers when the effective SSH parameters are tuned to topologically nontrivial values [Fig.~\ref{Fig_Floquet}a3]. These barrier states exhibit localized profiles in frequency space, as can be obtained by Fourier transforming the measured temporal voltage oscillations [Fig.~\ref{Fig_Floquet}a4]. Other than showcasing the utility of analog multipliers in effectively time-modulating circuit components at very high frequencies, this work (Ref.~\cite{stegmaier_topological_2024}) also demonstrated that Floquet circuit Laplacians can exhibit highly customizable inhomogeneities in the frequency lattice, an important realization that has far-reaching implications in topological signal processing.

Although time-varying components enable the construction of a frequency-domain lattice through their temporal oscillations, directly observing Floquet modes can be challenging due to the substantial heating in rapidly time-modulated elements. For the goal of experimentally realizing (mathematical) Floquet behavior, one approach to circumvent this restriction is to directly simulate the frequency lattice in real space. An attempt along these lines is the experiment by Zhang et al.~\cite{zhang_Floquet_observation_2024}, which observed Floquet modes along temporal dislocations, in analogy to usual spatial dislocations that host topological modes. As shown in Fig.~\ref{Fig_Floquet}b1, a (logically) time-dependent 2D higher-order topological lattice is realized as a 3D stack of 2D layers, where the third dimension represents frequency. The time-dislocation is spatially implemented as a sudden change in the inter-layer connections between the left and right halves of the setup. These inter-layer couplings represent the (logical) driving protocol. As a result, Floquet $0$ and $\pi$ topological modes exist in the band gaps [Fig.~\ref{Fig_Floquet}b2]. Notably, these modes exist at the dislocation between the left and right halves of the 3D circuit setup [Fig.~\ref{Fig_Floquet}b3], even though the spatial connectivity within each 2D layer remains completely homogeneous across the dislocation. These Floquet $0$ and $\pi$ modes were directly measured through impedance measurements [Fig.~\ref{Fig_Floquet}b4].

Time-modulated electrical circuits can also be used to demonstrate dynamical phenomena such as Thouless pumping~\cite{thouless_quantization_1983}. Traditionally, the Thouless pump describes the adiabatic, quantized transport of charge across a bulk gap~\cite{niu_quantised_1984}, as can be deduced from spectral flow arguments. In the regime of classical electrical circuits, charge quantization is not accessible, but spectral flow can still be measured since the Laplacian spectrum corresponds to impedance resonances. Stegmaier et al.~\cite{stegmaier_realizing_2024} demonstrated such spectral pumping in an electrical circuit with inductors that are effectively time-modulated with analog multipliers [Fig.~\ref{Fig_Floquet}c1]. As in Ref.~\cite{stegmaier_topological_2024}, the time modulation is controlled by an external voltage signal which can be customized at will---here, the modulation is chosen to yield the Aubry-André-Harper (AAH) model. By measuring the impedances within the circuit, eigenfrequency bands were reconstructed as a function of the periodically pumped phase. Indeed, the expected AAH spectral flow was observed under PBCs [Fig.~\ref{Fig_Floquet}c2], and topological in-gap states can also be seen under OBCs [Fig.~\ref{Fig_Floquet}c3]. It is important to note that, unlike the quantized charge transport predicted by the Kubo formula in quantum settings, here there is no occupied ground state, and the observed state pumping is purely that of the spectral band resonances.

Even in a static electrical circuit, techniques used to resolve Floquet dynamics can also be used to measure the time-dependent behavior. In the non-Hermitian corner localization experiment by Wu et al.~\cite{wu_evidencing_2023}, the 2D circuit lattice, hosting non-Hermitian second-order topological phases, incorporates INICs to implement asymmetric intra-cell couplings [Fig.~\ref{Fig_Floquet}d1]. The modular design of the circuit, featuring detachable INIC modules, allows switching between Hermitian and non-Hermitian configurations [Fig.~\ref{Fig_Floquet}d2]. In the Hermitian configuration, a Gaussian pulse injected into the center of the circuit spreads across the entire lattice, as shown by the time-resolved voltage profile [Fig.~\ref{Fig_Floquet}d3]. In contrast, in the non-Hermitian setting, the pulse localizes at a corner of the circuit [Fig.~\ref{Fig_Floquet}d4]. These time-resolved voltage responses demonstrate how electrical circuits can serve as practical platforms for observing dynamical physical phenomena in real time (see also Ref.~\cite{wang_realization_2023_prb}).

Another noteworthy result in Ref.~\cite{wu_evidencing_2023} is the demonstration of the generalized Brillouin zone using the Laplace transform rather than the Fourier transform applied to the measured voltage distribution. The non-Hermiticity renders the wavevector complex, making it infeasible to obtain the Brillouin zone directly from the Fourier transform. However, Wu et al. utilized the Laplace transform method to effectively recover the generalized Brillouin zone, where different onsite admittances correspond to different contours of the band structure. Such site-resolved field distributions are highly accessible in topolectrical circuits, which is crucial for studying higher-dimensional lattices. For example, in another work~\cite{wu_observing_2024}, Wu et al. observed degeneracy conversions, including Weyl point-to-nodal line conversion and nodal line-to-entangled nodal line conversion.

\subsection{Hyperbolic lattices and their realizations in electrical circuits}
\label{sec:hyperbolic}

\begin{figure}[ht!]
	\centering
	\includegraphics[width=\linewidth]{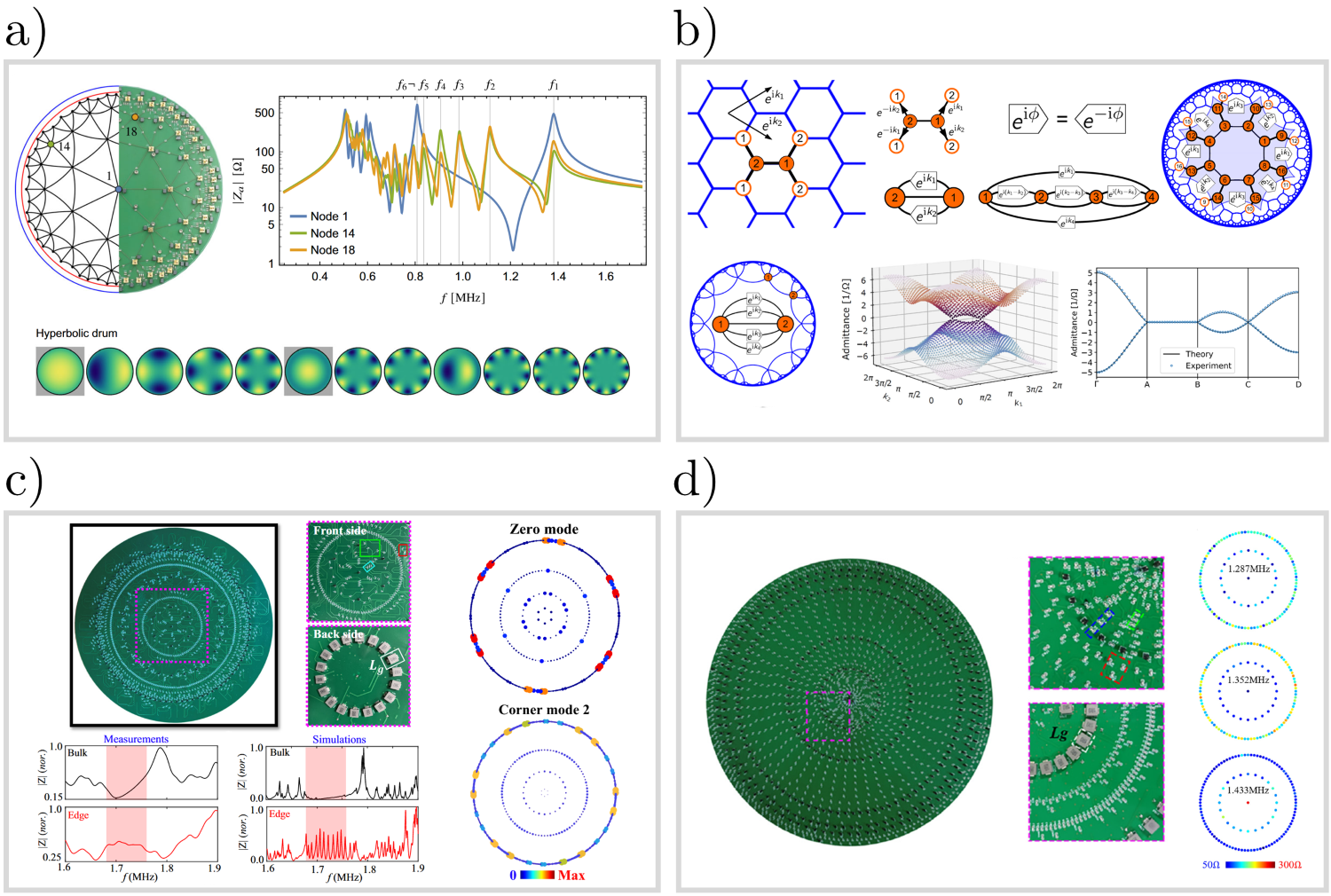}
	\caption{\textbf{Circuit implementations of hyperbolic lattices.} 
		\textbf{(a)} Hyperbolic ``drum''-shaped eigenmodes on a Poincaré disk, resolved through the phase information in the node voltages~\cite{lenggenhager_simulating_2022}. [Adapted with permission from Lenggenhager et al., Nat Commun 13, 4373 (2022). Copyright 2022 licensed under CC BY 4.0.]
		\textbf{(b)} Experimental simulation of a synthetic hyperbolic lattice with four effective hyperbolic band momenta. The synthetic momenta are implemented through adjustable phases, leading to the observation of graphene-like Dirac cones~\cite{chen_hyperbolic_2023}. [Adapted with permission from Chen et al., Nat Commun 14, 622 (2023). Copyright 2023 licensed under CC BY 4.0.]
		\textbf{(c)} Circuit realization of a Haldane hyperbolic lattice, with the number of measured topological zero modes increasing exponentially with lattice size~\cite{zhang_observation_hyperbolic_2022}. [Adapted with permission from Zhang et al., Nat Commun 13, 2937 (2022). Copyright 2022 licensed under CC BY 4.0.]
		\textbf{(d)} Demonstration of defect-induced backscattering-immune topological states on a hyperbolic lattice~\cite{pei_engineering_2023}. [Adapted with permission from Pei et al., Phys. Rev. B 107, 165145 (2023). Copyright 2023 by the American Physical Society.]}
	\label{Fig_Hyperbolic}
\end{figure} 

Lately, there has been much interest in the band structure and topological properties of hyperbolic lattices, which can be regarded as a vast generalization of regular lattices in negatively curved space~\cite{boettcher_crystallography_2022,urwyler_hyperbolic_2022,lv_hidden_2024,zhang_observation_hyperbolic_2022,tao_higher-order_2023,bzdusek_flat_2022,dey_simulating_2024,yuan_hyperbolic_2024,kollar_hyperbolic_2019,liu_chern_2022,yu_topological_2020,maciejko_hyperbolic_2021}. In 2D Euclidean (flat) space, 
regular $p$-gons can be tessellated such that every vertex contains $q$ branches, if $(p-2)(q-2)=4$. This leaves only three possibilities $\{p,q\}=\{3,6\},\{4,4\},\{6,3\}$ corresponding to the triangular, square and hexagonal lattices respectively. Relaxing this constraint to $(p-2)(q-2)>4$ results in an infinite variety of so-called hyperbolic lattices, which however need to be embedded in negatively curved space~\footnote{The opposite constraint $(p-2)(q-2)<4$ gives the five possibilities $\{p,q\}=\{3,5\},\{5,3\},\{3,4\},\{4,3\},\{3,3\}$,  respectively corresponding to the five platonic solids, i.e., icosahedron, dodecahedron, octahedron, cube and tetrahedron, all tiling the sphere.}. While this requirement of negatively curved space embedding makes them challenging to realize in most metamaterial platforms, electrical circuit connections can describe any desired graph structure, and thus emerge as promising physical platforms for hyperbolic lattices. This negative curvature fundamentally alters the propagation of modes, enriching the behavior of fundamental phenomena such as topological propagation~\cite{chen_anomalous_2024}. Due to the relative proliferation of boundary terminations, a hyperbolic lattice structure also alters the nature of the non-Hermitian bulk-boundary correspondences in the presence of asymmetric lattice hoppings~\cite{shen_non-hermitian_2025}. The ability of electrical circuits to simulate lattices with high coordination numbers is also helpful in the realization of higher dimensional lattices~\cite{lee_electromagnetic_2018,ezawa2018higher,ezawa2019non,zheng_exploring_2022,wang_circuit_2020,zhang_topolectrical-circuit_2020,zhang_exploring_2024,zheng_topological_2024}.

Lenggenhager et al.~\cite{lenggenhager_simulating_2022} achieved one of the first experimental demonstrations of hyperbolic lattices on a circuit board [Fig.~\ref{Fig_Hyperbolic}a]. By mapping the hyperbolic plane onto the Poincaré disk, the hyperbolic lattice takes the form of a tree with increasingly fine branches towards the boundary of the unit circle. Impedance resonances measurements reveal the eigenmodes of the hyperbolic lattice, which are distorted due to the negative curvature. By resolving the phase information at each node relative to a reference voltage, the structure of these ``hyperbolic drum'' eigenmodes can be mapped out.

To mitigate inherent difficulties in realizing large physical hyperbolic lattices, one strategy is to employ hyperbolic band theory~\cite{boettcher_crystallography_2022} which expresses a pseudo-infinite hyperbolic lattice in terms of hyperbolic translation operators. This was demonstrated by Chen et al.~\cite{chen_hyperbolic_2023}, who experimentally realized a graphene-like model in a pseudo-infinite synthetic hyperbolic lattice [Fig.~\ref{Fig_Hyperbolic}b] with just one physical unit cell. The synthetic parameter was implemented via complex-phase elements implemented with analog multipliers. The two-port Laplacian that introduces a phase is given by
\begin{equation}
	\begin{pmatrix}
		I_{1} \\ I_{2}
	\end{pmatrix} = \frac{1}{i\omega L} 
	\begin{pmatrix}
		1+i & e^{-i\phi} \\ e^{i\phi} &1+i 
	\end{pmatrix}
	\begin{pmatrix}
		V_1 \\ V_2
	\end{pmatrix},
\end{equation}
where $L$ is the inductance and the phase $\phi$ is controlled by adjusting the external voltages of the analog multipliers. The diagonal entries introduce a constant offset in the admittance spectrum, while the off-diagonal terms generate the controllable complex phase between ports $1$ and $2$.

A key feature of finite hyperbolic lattices is that the number boundary sites scales exponentially with the radius (i.e., number of lattice generations), reminiscent of fractal lattices~\cite{pai_topological_2019}. As such, unlike in Euclidean lattices, boundary sites can dominate the Hilbert space and greatly alter bulk-boundary correspondences, even in Hermitian setups. Zhang et al.~\cite{zhang_observation_hyperbolic_2022} demonstrated the exponential proliferation of zero modes in a  hyperbolic circuit [Fig.~\ref{Fig_Hyperbolic}c]. These topological zero modes give rise to a large number of impedance resonances at the edge nodes, while no resonances occur in the bulk nodes.

This boundary sensitivity is reflective of the small-world connectivity of a hyperbolic lattice, a property that has inspired their application in holographic duality~\cite{qi_exact_2013,gu2016holographic,lee2016exact,lee_generalized_2017,chen2023ads,brower_hyperbolic_2022,qi2018space}. The high connectivity also gives rise to heightened susceptibility to lattice defects or perturbations. Pei et al.~\cite{pei_engineering_2023} experimentally constructed a hyperbolic Haldane circuit and showed that a single defect in the bulk can induce boundary-dominated one-way propagation. By introducing polygonal bulk defects into the Haldane hyperbolic lattice, robust boundary propagation was observed, even when induced by a single point defect in the bulk [Fig.~\ref{Fig_Hyperbolic}d]. While the presence of zero modes was verified through impedance responses, the one-way propagation of topological edge states was demonstrated using dynamical voltage analysis.

\subsection{Simulations of other topological and quantum systems}
Electrical circuits are also well suited for simulating quantum media, both of conventional topological materials and more unconventional setups such as~\cite{wu_observing_2024,biao_realization_2023,bao_circuit_2023,rontgen_topological_2024}. Described below are some experimental demonstrations, particularly those of models that are hard to simulate on other platforms.

\begin{figure*}
	\centering
	\includegraphics[width=\textwidth]{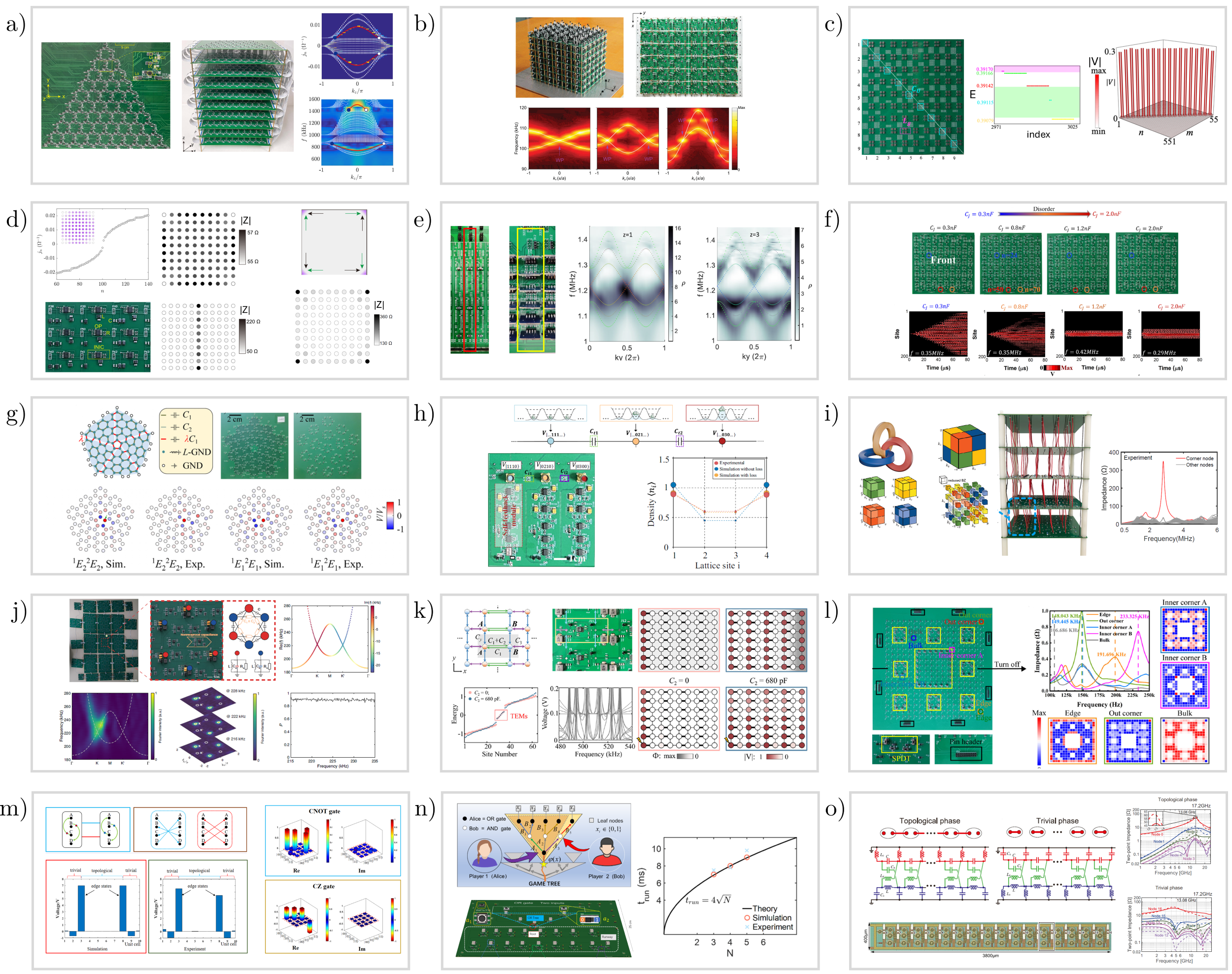}
	\caption{\textbf{Electrical circuit realizations of topological and quantum systems. }
		\textbf{(a)} Square-root topological lattice experiment showcasing higher-order Weyl semi-metallic phases~\cite{song_square-root_2022}. [Adapted with permission from Song et al., Nat Commun 13, 5601 (2022). Copyright 2022 licensed under CC BY 4.0.]
		\textbf{(b)} Type-II Weyl points in a 3D circuit with tilted bands~\cite{li_ideal_2021}. [Adapted with permission from Li et al., National Science Review, 8(7), nwaa192 (2021). Copyright 2021 licensed under CC BY 4.0.]
		\textbf{(c)} Interaction-induced two-boson flat-band localization and topological edge states simulated on a 2D circuit array~\cite{zhou_observation_2023}. [Adapted with permission from Zhou et al., Phys. Rev. B 107, 035152 (2023). Copyright 2023 by the American Physical Society.] 
		\textbf{(d)} Simulation of Wilson fermions in an electrical circuit, with the measurement of chiral edge currents along the domain walls of the associated Chern lattice model~\cite{yang_realization_2023}. [Adapted with permission from Yang et al., Commun Phys 6, 211 (2023). Copyright 2023 licensed under CC BY 4.0.]
		\textbf{(e)} 3D circuit realization and measurement of the Hopf insulator model~\cite{wang_realization_2023}. [Adapted with permission from Wang et al., Phys. Rev. Lett. 130, 057201 (2023). Copyright 2023 by the American Physical Society.]
		\textbf{(f)} Observation of non-Abelian Anderson localization and transitions through site-resolved impedance spectra and time-domain voltage dynamics~\cite{wang_observation_2023}. [Adapted with permission from Wang et al., Phys. Rev. B 108, 144203 (2023). Copyright 2023 by the American Physical Society.]
		\textbf{(g)} Demonstration of topological states and their spectral flows at disclinations, showcasing the correspondence between real-space topology and disclination states~\cite{xie_correspondence_2023}. [Adapted with permission from Xie et al., Phys. Rev. B 108, 134118 (2023). Copyright 2023 by the American Physical Society.] 
		\textbf{(h)} Simulation of many-body bound states in the continuum and their boundary-localized behavior in a non-linear circuit network~\cite{sun_boundary-localized_2024}. [Adapted with permission from Sun et al., Commun Phys 7, 299 (2024). Copyright 2024 licensed under CC BY 4.0.]
		\textbf{(i)} Circuit realization of the octupole higher-order topological insulating model and the observation of their corner states~\cite{qiu_octupole_2024}. [Adapted with permission from Qiu et al., Natl. Sci. Rev. nwaf137 (2025). Copyright 2025 licensed under CC BY 4.0.]
		\textbf{(j)} Non-Hermitian Haldane circuit exhibiting valley-dependent Dirac cones~\cite{xie_non-hermitian_2025}. [Adapted with permission from Xie et al., Nat Commun 16, 1627 (2025). Copyright 2025 licensed under CC BY 4.0.]
		\textbf{(k)} Circuit realization of topological states in 1D, 2D, and 3D that are delocalized by the NHSE~\cite{lin_evolution_2024}. [Reprinted from Lin et al., Appl. Phys. Lett. 125, 173104 (2024), with the permission of AIP Publishing.]
		\textbf{(l)} Fractal circuits with a proliferation of topological boundary modes~\cite{he_mode_2024}. [Adapted with permission from He et al., Phys. Rev. B 109, 235406 (2024). Copyright 2024 by the American Physical Society.]
		\textbf{(m)} Emulation of quantum braiding operations using classical circuits~\cite{zou_experimental_2023}. [Adapted with permission from Zou et al., Advanced Intelligent Systems, 5(11), 2300354 (2023). Copyright 2023 licensed under CC BY 4.0.]
		\textbf{(n)} Emulation of a quantum two-player game through a classical circuit~\cite{zhang_quantum_2024}. [Adapted with permission from Zhang et al., Research, 7, 0480 (2024). Copyright 2024 licensed under CC BY 4.0.]  
		\textbf{(o)} Integrated circuit (IC) realization of a topological Kitaev lattice with very high operating frequencies~\cite{iizuka_experimental_2023}. [Adapted with permission from Iizuka et al., Commun Phys 6, 279 (2023). Copyright 2023 licensed under CC BY 4.0.]}
	\label{Fig_OtherCM}
\end{figure*} 

One class of models that are particularly amenable to electrical circuit realizations are the square-root models. Given a parent Hamiltonian, the square-root of it is a lattice model in which two successive hoppings yield the same connectivity structure as the parent model (up to an unimportant constant shift)~\cite{song_realization_2020,sahin2021nth,yan_acoustic_2020,kremer_square-root_2020,ezawa_systematic_2020,marques_2n-root_2021,zhou_qth-root_2022,guo_realization_2023,wei_realization_2023,marques_generalized_2022,zhang_complexreal_2022,deng_n_2022,guo_multiple_2025,luo_exploring_2025}. Mathematically, they may look contrived and complicated, but they inherit various properties from their parent Hamiltonians, such as midgap topological modes~\cite{arkinstall_topological_2017}. The number of modes, including topological modes, is doubled in these lattices~\cite{marques_one-dimensional_2021}. Electrical circuits are suitable physical platforms due to their versatile connectivity, and Song et al.~\cite{song_square-root_2022} constructed a 3D square-root higher-order Weyl semi-metal by stacking 2D square-root lattices along the third dimension [Fig.~\ref{Fig_OtherCM}a]. From the admittance spectrum, non-zero mid-gap topological hinge states were observed.

Electrical circuits are also useful in demonstrating unconventional band structures, particularly in higher dimensions where fabrication may be challenging for some other metamaterials. For instance, Weyl points are crossings in 3D band structures, and can tilt over to form type-II Weyl points~\cite{hasan_weyl_2021,yan_topological_2017,wang_mote_2016,koepernik_tairte_2016,lin_line_2017,lu_probing_2019,zheng_exploring_2022,rafi-ul-islam_chiral_2024,biao_realization_2023,rafi-ul-islam_topoelectrical_2020,rafi-ul-islam_anti-klein_2020}, exhibiting great sensitivity during the so-called Lifshitz transition. Li et al.~\cite{li_ideal_2021} experimentally demonstrated type-II Weyl points in their 3D circuit [Fig.~\ref{Fig_OtherCM}b]. By exciting bulk states and measuring node voltages, topological surface states were measured, exhibiting tilted bands with two group velocities in the same direction---key properties of type-II Weyl systems~\cite{soluyanov_type-ii_2015}.

Beyond band structures, 2D or 3D electrical circuit arrays can also simulate few-body physics and other phenomena. By mapping the 2-photon degrees of freedom onto a 2D lattice~\cite{olekhno_topological_2020,lee_many-body_2021,jiao_two-dimensional_2021}, Zhou et al.~\cite{zhou_observation_2023} simulated interaction-induced flat-band localizations and topological edge states, tuning the effective interaction strength through circuit grounding. By measuring impedance responses and voltage dynamics in the time domain, they were able to resolve two-boson flat bands and topological edge states [Fig.~\ref{Fig_OtherCM}c]. On a different note, Yang et al.~\cite{yang_realization_2023} simulated Wilson fermions---a lattice discretization conventionally used in the context of lattice QCD to avoid fermion doubling---in a circuit lattice, tuning circuit parameters to achieve different Wilson fermion states. Through impedance measurements, they reconstructed the chiral edge domain wall separating two circuits with contrasting fractional Chern numbers, and observed localized corner states arising from two Chern lattices with opposite chiralities [Fig.~\ref{Fig_OtherCM}d]. Wang \textit{et al.}~\cite{wang_realization_2023} realized the elusive Hopf insulator in a 3D electrical circuit systematically implemented with so-called pseudospin modules. Although this lattice required long-range spin-orbit couplings, the pseudospin modules aided the realization of Hopf bands, including both surface and bulk bands [Fig.~\ref{Fig_OtherCM}e].

Electrical circuits have also been used to simulate non-Abelian systems, in which physical outcomes depend on the sequence of operations, i.e., twists~\cite{nayak_non-abelian_2008,bouhon_non-abelian_2020,bouhon_wilson_2019,yang_non-abelian_2024,wu_non-abelian_2019,tang_experimental_2022}. Various recent experiments have attempted to classically simulate operations that are mathematically equivalent to that experienced by anyonic quasiparticles~\cite{zhang_observation-Bloch_2022,chen_classical_2022,zhang_anyonic_2023,zhou_bloch_2024,zhou_anyonic_2025,qin_dynamical_2025}. Wang et al.~\cite{wang_observation_2023} demonstrated non-Abelian Anderson localization using a 2D electrical circuit that represents the quasiperiodic AAH model with non-Abelian gauge fields. Both localization and delocalization behaviors were observed through site-resolved impedance spectra and voltage dynamics [Fig.~\ref{Fig_OtherCM}f] (see also the experimental realization of non-Abelian inverse Anderson transition in Ref.~\cite{zhang_non-abelian_2023}). 

Circuits are also particularly suitable for demonstrating the effects of lattice defects, since nodes are connected at will. Xie et al.~\cite{xie_correspondence_2023} experimentally demonstrated disclination states and their correspondence to real-space topology. Localized topological states at the disclinations---defects in rotational symmetry~\cite{teo_topological_2017}---were observed in impedance maps and voltage distributions at the lattice nodes [Fig.~\ref{Fig_OtherCM}g] (see also Refs.~\cite{xu_topolectrical_2025,li_observation_2025,liu_experimental_2025}).

The notion of bound states in the continuum (BICs)~\cite{hsu_bound_2016}, which are localized and isolated from the environment, is rapidly extending beyond photonics~\cite{marinica_bound_2008,kang_applications_2023,nasari_non-hermitian_2023}, ring resonators~\cite{guo_observation_2021} and into electrical circuits~\cite{zhang_anyonic_2023,luo_exploring_2025,wang_exotic_2024}. Sun et al.~\cite{sun_boundary-localized_2024} simulated boundary-localized many-body BICs by mapping the Fock states of three bosons into non-linear circuit networks. The three circuit nodes corresponding to the three bosonic states are connected by capacitors, and each node consists of a self-feedback module controlling the effective interaction strength. These modules, which involve op-amps and analog multipliers, allow for the simulation of few-boson BICs with different effective interaction strengths [Fig.~\ref{Fig_OtherCM}h].

Blessed with the freedom in implementing any desired boundary condition, electrical circuit arrays can assume unconventional boundary conditions not possible in materials and most metamaterials, such as higher-dimensional real projective planes or even non-orientable manifolds, with the exception of some photonic systems~\cite{wang_experimental_2023}. Combined with nontrivial momentum-space higher-order topology arising from the octupole moment of the bulk, Qiu et al.~\cite{qiu_octupole_2024} experimentally demonstrated a 3D lattice [Fig.~\ref{Fig_OtherCM}i] which exhibits defect-like higher-order corner states through impedance analysis. The versatility of electrical circuit has enabled the realization of such unconventional states resulting from the interplay of non-trivial real-space and momentum-space topology~\cite{xie_higher-order_2021,bao_topoelectrical_2019}.

Since electrical circuits commonly involve resistive and active elements, they can readily demonstrate how non-Hermiticity interplays with topology without much additional complication. Xie et al.~\cite{xie_non-hermitian_2025} demonstrated the non-Hermitian version of the graphene lattice, which exhibits two non-Hermitian Dirac cones, one experiencing amplification and the other experiencing decay. Site-resolved band measurements reveal unidirectional kink states~\cite{rafi-ul-islam_valley_2023,zhang_effects_2023} with dissimilar profiles at the two valleys [Fig.~\ref{Fig_OtherCM}j]. Involving asymmetrical real next-nearest-neighbor couplings effectively results in a non-Hermitian single Dirac cone, leading to valley polarization~\cite{behnia_polarized_2012,rafi-ul-islam_valley_2022,rafi-ul-islam_conductance_2023,siu_valley_2023} with a large bandwidth [see also Ref.~\cite{li_realization_2025}].

The introduction of non-Hermiticity can also greatly alter the locality of topological modes. While topological modes are typically strongly confined to the boundaries~\cite{hasan_colloquium_2010}, non-Hermitian skin pumping in the opposite direction can delocalize them while preserving their eigenspectrum, in this case mid-gap zero-eigenvalues~\cite{wang_non-hermitian_2022}. These delocalized topological states, still protected by lattice symmetries, can extend through the bulk, enabling their robustness to be harnessed not only at the boundaries but also within the bulk~\cite{cai_nonlinearity-driven_2024,bai_nonlinearity-enabled_2023}. Lin et al.~\cite{lin_evolution_2024} demonstrated these extended states in 1D, 2D, and 3D circuits by tuning the non-reciprocal coupling strength [Fig.~\ref{Fig_OtherCM}k]. 
At a critical strength determined by both the topological and skin localization length, the topological modes become extended, a phenomenon that was also studied in Refs.~\cite{lu_extended_2024,liu_extended_2024,ferdous_observation_2023,ruan_interlayer_2024,li_localized_2024}.

Beyond hyperbolic lattices (refer to Section~\ref{sec:hyperbolic}), electrical circuits can also simulate more esoteric network structures, such as fractals. On such networks, boundary terminations occur at various length scales, and topological modes localized unconventionally in the form of extended bulk states and inner states~\cite{manna_inner_2023,sun_inner_2025,sun_non-hermitian_2024,pai_topological_2019,bai_arbitrarily_2024}. He et al.~\cite{he_mode_2024} demonstrated fractal topological circuits which exhibit a rich spectrum of topological edge and corner modes. Reminiscent of hyperbolic lattices, the numbers of such boundary modes increase rapidly with system size, as directly observed through impedance analysis [Fig.~\ref{Fig_OtherCM}l].

Topolectrical circuits can simulate not only classical systems, but also quantum setups~\cite{ji_fast_2022,pan_electric-circuit_2021,zhang_new_2023}. By mapping the Hilbert space of few-qubit systems to the nodes of suitably designed circuits, Zou et al.~\cite{zou_experimental_2023} demonstrated [Fig.~\ref{Fig_OtherCM}m] the mathematical equivalences of Majorana-like edge states, T-junctions, two-qubit unitary operations, and Grover’s search algorithm. Through their resistor-capacitor-based circuit with segmented fixed resistances, braiding processes can be simulated without significant loss. Zhang et al.~\cite{zhang_quantum_2024} demonstrated another example of quantum algorithm simulation using classical circuits. They simulated a two-player quantum game based on AND-OR tree structures [Fig.~\ref{Fig_OtherCM}n]. Relays were employed to disconnect each node simultaneously after applying the initial voltage, allowing the system to evolve from the initial state. 

Although most topolectrical circuit simulations of condensed matter have been built on printed circuit boards, a few experiments have demonstrated their implementation in integrated circuits (ICs). For instance, Iizuka et al.~\cite{iizuka_experimental_2023} experimentally demonstrated topological Kitaev interface states~\cite{liu_experimental_2023} in high-frequency ICs [Fig.~\ref{Fig_OtherCM}o]. Through two-point impedance measurements, this IC realization successfully captured both topological and trivial phases at frequencies around 13 GHz. These IC realizations, with their much higher degree of miniaturization, demonstrate the great potential for topolectrical circuits in scalable high-frequency applications.

\begin{figure}[ht!]
	\centering
	\includegraphics[width=\linewidth]{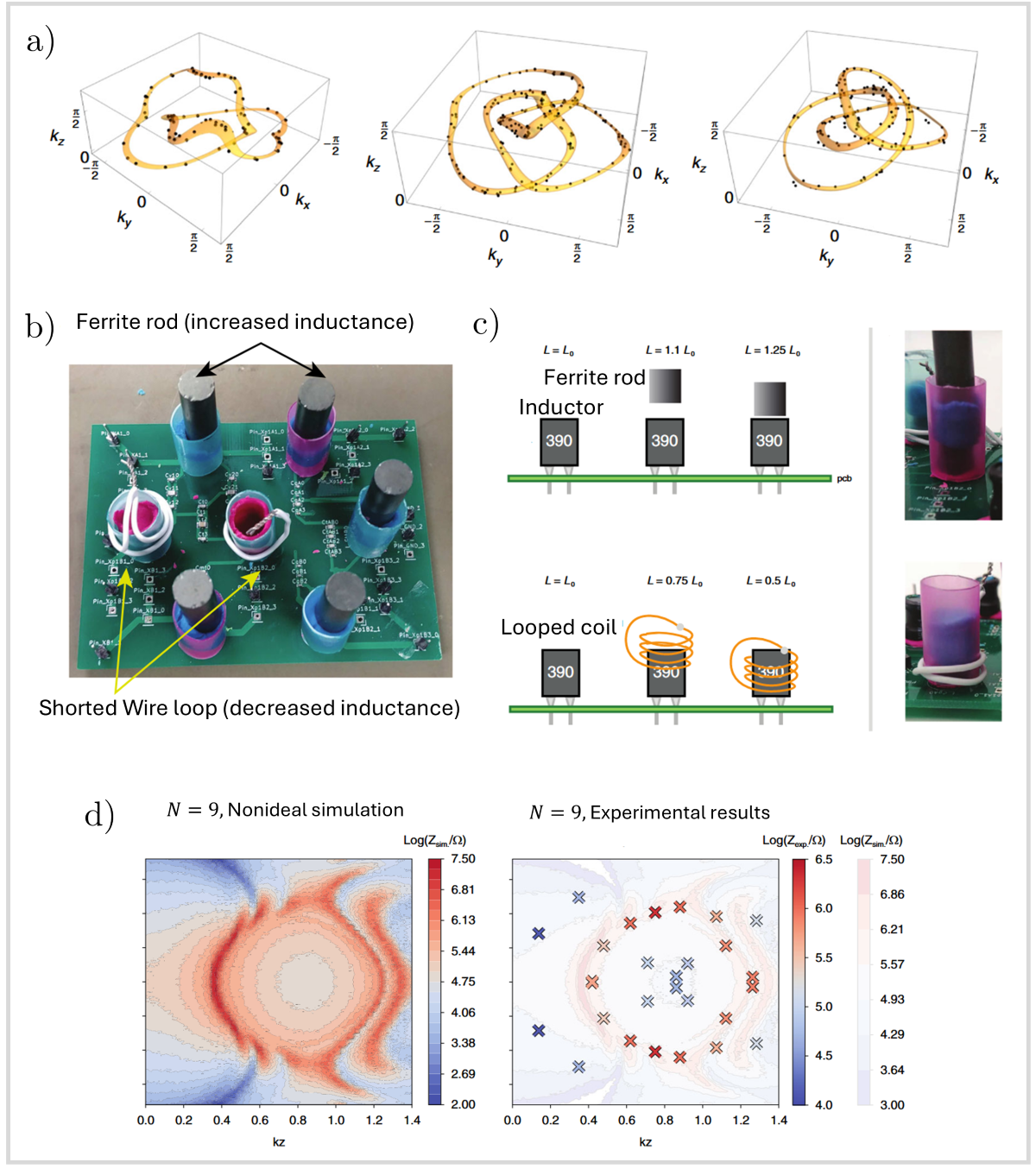}
	\caption{\textbf{Imaging nodal knots through electrical circuits. }  
		\textbf{(a)} Nodal knot construction from the method in Ref.~\cite{lee_imaging_2020} [Eqs.~\eqref{J1} and \eqref{J2}]: Theoretical nodal links or knots shown in orange, with simulated admittance eigenvalues represented as black points in 3D momentum space.  
		\textbf{(b)} A unit cell of the experimental circuit featuring variable inductors (cylinders).  
		\textbf{(c)} Mechanism behind the variable inductors: ferrite rods increase the inductance and a shorted wire loop decreases it, by up to 25\% and 50\% respectively.  
		\textbf{(d)} The simulated drumhead region impedance of the Hopf link (left) agrees well with measurements at optimally chosen momentum points (right). [(a,b,c,d) adapted with permission from Lee et al., Nat Commun 11, 4385 (2020). Copyright 2020 licensed under CC BY 4.0.]}
	\label{Fig_NodalKnots}
\end{figure}

\subsubsection{Nodal knots and topological drumhead surface states}

While various experiments have simulated knotted band touchings in 3D parameter space, in both Hermitian and non-Hermitian (exceptional knots) contexts~\cite{zhang_exploring_2024,cao_observation_2024,zhang_experimental_2023}, realizing these knots in momentum space has the benefit of accessing their topological surface states at real-space terminations. 
Electrical circuits prove particularly suitable due to the ease in implementing long-range couplings, which are essential for the construction of nodal knots in momentum space~\cite{bian_drumhead_2016,bi_nodal-knot_2017,li_emergence_2019,luo_topological_2018,tai2021anisotropic,ding_non-hermitian_2022,zhang_tidal_2021,rafi-ul-islam_knots_2024,rafi-ul-islam_twisted_2024,burkov_topological_2011}. In general, the topological surface states of nodal knots lie within the interior of the 2-dimensional surface Brillouin zone projections of these knots, and are thus known as ``drumhead'' surface states~\cite{luo_topological_2018}.

Below, following Ref.~\cite{lee_imaging_2020}, we describe the construction of nodal knot models that are suitable for platforms with reciprocal elements, i.e., passive RLC circuits. In a reciprocal lattice with two nodes per unit cell, a minimal circuit Laplacian ansatz can be written as
\begin{equation}
	J(\mathbf{k})=l_0 \mathbb{I}+\mathfrak{M} \mathrm{e} f(\mathbf{k}) \sigma_x+\mathfrak{I m} f(\mathbf{k}) \sigma_z,
	\label{J1}
\end{equation}
where $l_0$ imposes a uniform offset, $f(\mathbf{k})$ is a complex function of the lattice momentum, and $\sigma_x$ and $\sigma_z$ are Pauli matrices. The bands cross in the 3D Brillouin zone where $2|f(\mathbf{k})|=0$, i.e., when the coefficients of both Pauli matrices disappear. To facilitate the knot construction, $\mathbf k\in \mathbb{T}^3$ is first mapped onto two complex numbers $(z,w)$:
\begin{equation}
	\begin{aligned}
		&z(\mathbf k) =\cos 2 k_z+\frac{1}{2}+i\left(\cos k_x+\cos k_y+\cos k_z-2\right), \\
		&w(\mathbf k) =\sin k_x+i \sin k_y,
		\label{J2}
	\end{aligned}
\end{equation}
such that different knots correspond to different bivariate polynomials in $z,w$: for instance, the $(p,q)$-torus knots correspond to $f(\mathbf k) = z(\mathbf k)^p + w(\mathbf k)^q$~\cite{lee_imaging_2020}. Such polynomials were already devised for various knots~\cite{bode_constructing_2019,dennis_isolated_2010,bode_knotted_2017}, but their embedding onto the 3D Brillouin zone, where $z(\mathbf k),w(\mathbf k)$ are both periodic functions of $\mathbf k$, has only been considered in the topological nodal context. 

As simulated and experimentally demonstrated in Ref.~\cite{lee_imaging_2020}, nodal structures can be resolved from the admittance eigenvalues [Fig.~\ref{Fig_NodalKnots}a]. These nodal knots form intricate patterns in the 3D momentum space, while the corresponding drumhead surface states appear as projected flat bands in the surface admittance spectrum. While the edge-terminated direction was implemented in real space, the two other directions in the plane of the drumhead surface states were parametrically tuned by adjusting the inductance of the inductors to correspond to different momentum values [Fig.~\ref{Fig_NodalKnots}b]. The inductance was varied using ferrite rods or shorted wire loops to increase or decrease inductance, respectively [Fig.~\ref{Fig_NodalKnots}c], effectively reducing the undesired effects of parasitics and component uncertainties. By utilizing machine learning-powered optimization and microcontrollers to fine-tune the setup, the drumhead regions could be resolved with a minimal number of measurements [Fig.~\ref{Fig_NodalKnots}d].

\section{Machine learning approaches to topolectrical circuit construction and measurement}

\begin{figure}[ht!]
	\centering
	\includegraphics[width=\linewidth]{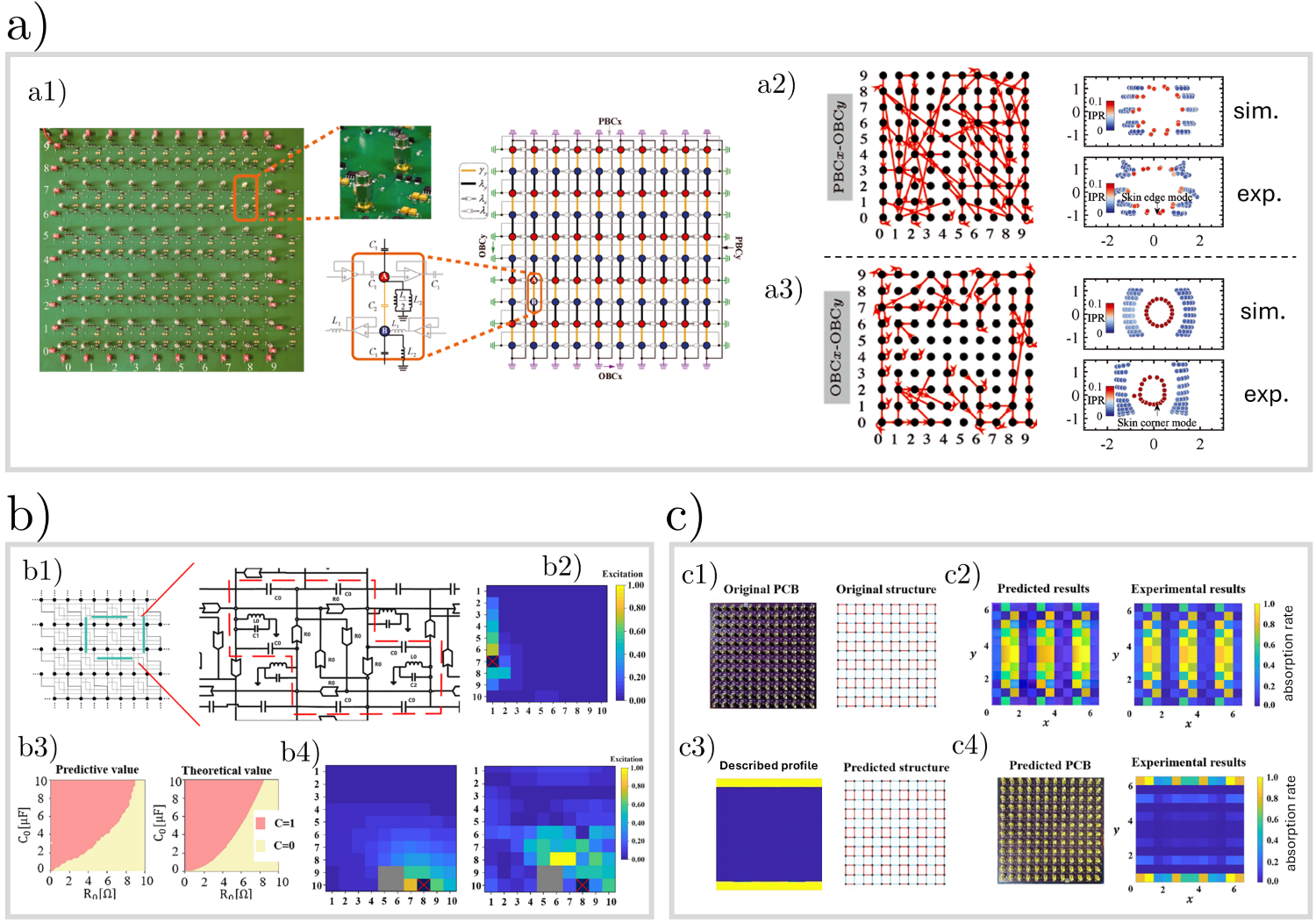}
	\caption{\textbf{Applying machine learning to topolectrical circuit experiments.} 
		\textbf{(a)} Application of the physics-graph-informed machine learning approach in streamlining the measurement of the second-order non-Hermitian skin effect in a 2D circuit, and the determination of its eigenspectrum~\cite{shang_experimental_2022}. [Adapted with permission from Shang et al., Advanced Science, 9(36), 2202922 (2022). Copyright 2022 licensed under CC BY 4.0.]
		\textbf{(b)} Application of deep learning to the design of the topological model for a circuit realization~\cite{chen_deep_2024}. [Adapted with permission from Chen et al., Phys. Rev. B 109, 094103 (2024). Copyright 2024 by the American Physical Society.]
		\textbf{(c)} Design of electrical circuits based on text or image inputs, experimentally tested with a 2D topological circuit~\cite{chen_composable_2024}. [Adapted with permission from Chen et al., Phys. Rev. B 110, 134108 (2024). Copyright 2024 by the American Physical Society.]}
	\label{Fig_ML}
\end{figure} 

Topolectrical circuit simulation of a condensed matter lattice usually entails the design, construction and voltage measurement of a large number of nodes. To streamline the process, some studies have employed machine learning (ML) approaches, in particular deep learning, for optimizing their design and measurement process. In other contexts, ML approaches have already proven themselves as powerful tools for optimization and prediction based on incomplete information, for instance extrapolating the dynamics of complex systems as well as optimizing their design~\cite{gilpin_generative_2024,panahi_machine_2024,noe_machine_2020}. As such, they are particularly effective in addressing practical challenges such as managing a large number of circuit variables, designing printed circuit boards with minimal effective node-to-node connections, and streamlining circuit measurements.

To illustrate, fully acquiring the Laplacian spectrum of a circuit requires $N \times N$ impedance measurements in principle, where $N$ is the total number of nodes. For large $N$, this process becomes increasingly tedious, especially in higher-dimensional circuits. To address this, Shang et al.~\cite{shang_experimental_2022} introduced the physics-graph-informed machine learning approach for reconstructing the circuit spectrum with incomplete data [Fig.~\ref{Fig_ML}a1], towards a similar objective as Ref.~\cite{franca_impedance_2024}. Using physics-graph-informed ML, they conclusively demonstrated second-order NHSE spectra through impedance measurements only between selected ports [Figs.~\ref{Fig_ML}a2 and a3]. Aided by the $K$-means clusters algorithm, only $N^2/K$ measurements were required, highlighting how embedding physics-informed priors into ML frameworks can significantly enhance data interpretation and analysis in topolectrical circuit experiments.

In addition to data analysis, deep learning can aid the design of the mathematical model itself. Chen et al.~\cite{chen_deep_2024} utilized multilayer perceptrons and convolutional neural networks to predict topological invariants and the circuit response. They simulated the topolectrical Chern circuit~\cite{hofmann_chiral_2019} [Fig.~\ref{Fig_ML}b1] with 1000 different sets of circuit component parameters to train their model. The topolectrical Chern circuit features chiral voltage edge propagation when a Gaussian pulse is injected at an edge node [Fig.~\ref{Fig_ML}b2]. Their trained model demonstrated the approximate phase diagram [Fig.~\ref{Fig_ML}b3] and captured aspects of the temporal evolution of chiral voltage propagation, even in the presence of defects [Fig.~\ref{Fig_ML}b4].

Chen et al.~\cite{chen_composable_2024} proposed a framework that integrates text, image, and spectral data to assist in the design of two-dimensional SSH circuits. Their bidirectional collaborative design framework enables both forward prediction of measured quantities from the described circuit structure and inverse design of topolectrical circuits based on textual descriptions. For instance, when provided with an image of a fabricated 2D SSH circuit [Fig.~\ref{Fig_ML}c1], the trained model inferred circuit properties that aligned with experimental measurements [Fig.~\ref{Fig_ML}c2]. Conversely, given a target voltage profile (e.g., edge-localized voltage [Fig.~\ref{Fig_ML}c3]), the model suggested a corresponding circuit design [Fig.~\ref{Fig_ML}c4]. The fabricated circuit based on this predicted structure exhibited a voltage profile similar to the intended design.

\section{Outlook}

Electrical circuits, utilizing components such as conventional elements, varactors, op-amps, analog multipliers, and memristors~\cite{di_ventra_custodial_2022,strukov_missing_2008}, enable the realization of features and phenomena that are hard to simulate in most other platforms, such as higher-order topological phases~\cite{ezawa_electric_2019,guo_realization_2024}, Floquet setups~\cite{zhang_Floquet_observation_2024,stegmaier_topological_2024}, hyperbolic lattices~\cite{zhang_hyperbolic_2023}, and synthetic dimensions~\cite{yu_4d_2020}. Recent advances have expanded the scope of electrical circuit experiments to include the simulation of flat bands~\cite{banerjee_non-hermitian_2024,biao_experimental_2024}, interaction-induced relativistic effects~\cite{zhang_observation_2021}, quantum valley Hall effects~\cite{zhu_quantum_2019}, exceptional Landau quantization~\cite{zhang_non-hermitian_2020}, Hopf bundles~\cite{kim_realization_2023}, inverse Anderson transitions~\cite{wang_observation_2022}, and non-Abelian topological bound states~\cite{qian_non-abelian_2024}, offering unprecedented control and tunability. These demonstrations highlight the adaptability of topolectrical circuits in bridging theoretical models with physical realizations. Their direct observability and scalability present unique opportunities for uncovering new physics. Furthermore, machine learning and deep learning have enhanced the potential of electrical circuits, enabling their efficient analysis, prediction, and design, even in systems with complicated topological features or incomplete data~\cite{shang_experimental_2022}.

The integration of electrical circuits with advanced computational techniques and emerging technologies~\cite{ni_robust_2020} promises to deepen our understanding of condensed matter phenomena and push the boundaries of experimental physics~\cite{yang_circuit_2024,chen_engineering_2025,shen2025observation}. Their adaptability~\cite{yatsugi_observation_2022,lee_imaging_2020} to simulate higher-dimensional systems~\cite{wang_circuit_2020}, such as those seen in magic-angle Moiré circuits~\cite{zhang_moire_2021}, opens new avenues for investigating uncharted areas, including relatively esoteric condensed matter systems~\cite{sun2024self,jiang_four-band_2021,ni_higher-order_2021,wang_berry_2024,huang_topological_2024,pan_nested_2025,li_multichannel_2024}.

Potential future directions in topolectrical circuits include the integration of tunable active elements, such as digitally controlled inductors and analog multipliers, enabling real-time parameter adjustments for Floquet circuit implementations. Although not yet widely available commercially, memristors---often considered the fourth passive circuit element~\cite{chua_memristor-missing_1971,strukov_missing_2008,jain_heterogeneous_2025} alongside capacitors, inductors, and resistors---could facilitate memory-based physical phenomena such as axon-like transmissions~\cite{brown_axon-like_2024} or protected topological states even with a broken symmetry~\cite{di_ventra_custodial_2022}. The incorporation of other nonlinear elements, such as transistors and various types of diodes, remains an open avenue that could simplify circuit implementations and broaden potential applications. These advancements may transition topolectrical circuits from proof-of-concept research to real-world applications in signal processing, computation, and sensing. Further refinement and expansion of the scope of topolectrical circuits could lead to uncovering fundamental insights and advance the development of practical applications in condensed matter and beyond.

\section*{Acknowledgments}
This work is supported by the Ministry of Education (MOE) Tier-II Grants No. MOE-T2EP50121-0014 (NUS Grant No. A-8000086-01-00) and MOE-T2EP50222-0003 (NUS Grant. No. A-8001542-00-00), as well as MOE Tier-I FRC Grants (NUS Grant Nos. A-8000195-01-00 and A-8002656-00-00). \\

\section*{Author Declarations}
\subsection*{Conflict of Interest}
The authors declare no competing interests.
\subsection*{Data Availability}
Data sharing is not applicable to this article as no new data were created or analyzed in this study.

\bibliography{TEreview_arXiv_v2}

\end{document}